\documentclass[prd,nofootinbib,preprintnumbers,twocolumn,showpacs,floatfix,superscriptaddress]{revtex4-1}

\usepackage[utf8]{inputenc}      
\usepackage{amsmath}
\usepackage{amssymb}
\usepackage{graphicx}
\usepackage{color}
\usepackage{hyperref}

\newcommand{\bitem}{\begin{itemize}}
\newcommand{\eitem}{\end{itemize}}
\newcommand{\bwt}{\begin{widetext}}
\newcommand{\ewt}{\end{widetext}}
\newcommand{\beq}{\begin{equation}}
\newcommand{\eeq}{\end{equation}}

\newcommand\g{\gamma}
\newcommand{\bdm}{\begin{displaymath}}
\newcommand{\edm}{\end{displaymath}}
\newcommand{\bea}{\begin{eqnarray}}
\newcommand{\eea}{\end{eqnarray}}

\newcommand{\nn}{\nonumber}

\renewcommand\H{\mathcal H}
\renewcommand\P{\mathcal P}
\newcommand{\SM}{$\boldsymbol{\eps}'_{\bf SM}\ $}
\newcommand{\NP}{$\boldsymbol{\eps}'_{\bf NP}\ $}
\renewcommand{\a}{\alpha}
\renewcommand{\b}{\beta}
\renewcommand{\t}{\theta}
\def\C{\mathcal C}
\newcommand\e{\mathrm e}
\newcommand\sand[3]{\big\langle\overline{#1#2}\left|#3\right|{#1#2}\big\rangle}
\def\eq#1{{Eq.~(\ref{#1})}}
\def\eqs#1#2{{Eqs.~(\ref{#1})--(\ref{#2})}}

\def\fig#1{{Fig.~\ref{#1}}}

\def\Table#1{{Table~\ref{#1}}}

\def\sect#1{{Sect.~\ref{#1}}}

\def\app#1{{Appendix~\ref{#1}}}

\def\vev#1{\left\langle #1 \right\rangle}
\def\bra#1{\left\langle #1 \right|}
\def\ket#1{\left| #1 \right\rangle}

\def\Im{\mathrm{Im}\,}
\def\Re{\mathrm{Re}}

\def\det{\mbox{det}}

\newcommand\GeV{\text{GeV}}
\newcommand\TeV{\text{TeV}}
\newcommand\eps{\varepsilon}

\begin{document}
\jot = 1.4ex         

\title{Kaon CP violation and neutron EDM in the minimal left-right symmetric model}

\author{Stefano Bertolini}
\email{stefano.bertolini@sissa.it}
\affiliation{INFN, Sezione di Trieste, SISSA,
Via Bonomea 265, 34136 Trieste, Italy\\[1ex]}

\author{Alessio Maiezza}
\email{Alessio.Maiezza@irb.hr}
\affiliation{Rudjer Boskovic Institute, Division of Theoretical Physics, Bijenička cesta 54, 10000, Zagreb, Croatia\\[1ex]}

\author{Fabrizio Nesti}
\email{fabrizio.nesti@aquila.infn.it}
\affiliation{Dipartimento di Scienze Fisiche e Chimiche, Universit\`a dell'Aquila, via Vetoio SNC, I-67100, L'Aquila, Italy}


\begin{abstract} \noindent Within the minimal Left-Right (LR) symmetric model we revisit the
  predictions for the kaon CP violating observables $\eps$ and $\eps'$ in correlation with the
  neutron electric dipole moment. We perform a complete study of the cross constraints on the model
  parameters, phases and the $M_{W_R}$ scale, considering the two cases of extended parity or charge
  conjugation as LR discrete symmetries, together with the possible presence of a Peccei-Quinn
  symmetry. We discuss in particular two scenarios: whether the Standard Model saturates the
  experimental value of $\eps'/\eps$ or whether new physics is needed, still an open issue after the
  recent lattice results on the QCD penguin matrix elements. Within the first scenario, we find no
  constraints on the LR scale in the charge-conjugation case while in the parity case we show that
  $M_{W_R}$ can be as low as $13\,\TeV$. On the other side, the request that new physics contributes
  dominantly to $\eps'$ implies strong correlations among the model parameters, with an upper bound
  of $M_{W_R}< 8$--$100\,\TeV$ depending on $\tan\beta$ in the case of charge conjugation and a
  range of $M_{W_R}\simeq 7$--$45\,\TeV$ in the parity setup. Both scenarios may be probed directly
  at future colliders and only indirectly at the LHC.
\end{abstract}


\maketitle

\section{Introduction}

\noindent
Flavor phenomenology offers a window for physics beyond the Standard Model (SM). In particular,
flavor changing neutral current (FCNC) processes play a key role in the search for new phenomena
since they are forbidden at the tree level. For processes involving light quark families, on top of
the loop suppression a further reduction results from the smallness of the family quark mass
splittings, known as the GIM mechanism~\cite{Glashow:1970gm}. Moreover, CP violation requires the
presence of the three families in the loop and therefore of the hierarchically small
mixings~\cite{Kobayashi:1973fv}. The rarity of these processes is indeed a smashing success of the
SM setup. Kaon CP-violating (CPV) observables as $\eps$ and $\eps'$ belong to this class and are a
most sensitive probe for most extensions of the standard electroweak scenario.

A flavor conserving observable that shares a similar discovery potential is the electric dipole
moment of the neutron ($n$EDM)~\cite{Engel:2013lsa}.  It violates parity and time-reversal and
therefore CP. In the SM direct electroweak contributions are generated at higher loop order and are
well below the present experimental bound ($2.9\times 10^{-26}$ e cm~\cite{Baker:2006ts}). A direct
contribution related to the QCD theta-term $\bar\theta$ also induces an $n$EDM, which then requires
$\bar{\theta} < 10^{-10}$.

Such a tiny bound is technically natural in the SM because $\bar\theta$ is perturbatively
protected~\cite{Ellis:1978hq}, thus avoiding the need for a first principle understanding of its
smallness, the so-called strong CP problem.  On the other hand, the issue is real in most SM
extensions which exhibit new flavor structures and additional CPV phases that lead to potentially
large contributions to these observables.

In the present work, we update on the scrutiny of Left-Right (LR) symmetric theories, based on the
gauge group
$\mathcal{G}=SU(2)_L \otimes SU(2)_R \otimes
U(1)_{B-L}$~\cite{Pati:1974yy,Mohapatra:1974hk,Mohapatra:1974gc}. A particular role is played by
the minimal version of the Left-Right symmetric models
(LRSM)~\cite{Senjanovic:1975rk}. Besides being predictive, the model
provides a natural rationale for the origin and smallness of the
neutrino mass~\cite{Minkowski:1977sc,Mohapatra:1979ia}, a setup for the restoration of parity at high
scale~\cite{Senjanovic:1978ev}, and a novel source for neutrinoless double-beta
decay~\cite{Mohapatra:1980yp,Tello:2010am}.  The LRSM has aroused a renewed interest in the era of
LHC, because of the possibility of a direct detection via the Keung-Senjanovic (KS)
process~\cite{Keung:1983uu}, which violates lepton number in full analogy with the low energy
neutrinoless double-beta decay.

The minimality of the model ensures a connection between Majorana and Dirac masses, making it a
predictive theory even for
neutrinos~\cite{Nemevsek:2012iq,Senjanovic:2016vxw,Senjanovic:2018xtu,Helo:2018rll}.  Since early
times the LR scale $M_{W_R}$ as well as the model parameters were found to be strictly constrained
by flavor physics~\cite{Beall:1981ze,Ecker:1985vv}. More recently the possibility of a low LR scale
within the LHC reach was emphasized combining bounds from many observables~\cite{Maiezza:2010ic}.
Subsequent analyses focusing on $\eps'$ were performed~\cite{Bertolini:2012pu,Bertolini:2013noa} and
a lower bound on the LR scale slightly above 3\,TeV was finally evinced~\cite{Bertolini:2014sua},
including the relevant constraints from $B_{d,s}$ meson oscillations.

The impact of the $n$EDM bound on the LRSM requires a separate comment. Exact and spontaneously
broken $\P$ has been considered as a solution of the strong CP problem~\cite{Mohapatra:1978fy} as an
alternative to the dynamical Peccei-Quinn (PQ)
mechanism~\cite{Peccei:1977hh,Weinberg:1977ma,Wilczek:1977pj}.  In the case of spontaneously broken
$\P$, $\bar{\theta}$ becomes computable in terms of a single CP-violating parameter. The original
argument was revisited in~\cite{Maiezza:2014ala} uncovering a large bound on the LR scale of about
$M_{W_R}>30\,\TeV$, which pushes the scale beyond the current experimental reach. This conclusion is
however specific of the given setup and could be spoiled by envisaging, for instance, the presence
of a PQ mechanism so that the LR scale might still be at the reach of LHC. Although $M_{W_R}$ below
$6$\,TeV is disfavored by the demand that the Higgs sector remains in a perturbative
regime~\cite{Maiezza:2016bzp,Maiezza:2016ybz,Chauhan:2018uuy}, an energy window remains for
discovery via the KS process~\cite{Nemevsek:2018bbt}.

As we shall see, the phase and flavor structure of the LRSM tightly correlates $n$EDM, $\eps$ and
$\eps'$, calling for a comprehensive and detailed study. We shall not enter here the debate on the SM calculations of
$\eps'$ that presently suffer from large uncertainties and leave open the possibility of large new
physics contributions~\cite{Gisbert:2017vvj,Buras:2018ozh}.  Our analysis will address the different
scenarios according to the relevance of the LRSM contributions, addressing the present and future
implications.

Early detailed studies are found in~\cite{Frere:1990cj,Frere:1991jt,Frere:1991db}.
The topic has received a renewed interest in the last few years. 
The works~\cite{Cirigliano:2016yhc,Dekens:2017hyc} address the problem via an effective theory of RH
currents in a model independent way which necessarily misses the detailed phase correlations.
Ref.~\cite{Haba:2018byj} analyzes a specific choice of LR discrete symmetry (the left-right charge
conjugation $\C$, see below) and we shall compare their findings with our results.  In particular we
point out an issue in the calculation of the relevant meson-baryon couplings that affects the calculation of the chiral loop contributions and alters
substantially the conclusions  in the PQ scenario.  In \cite{Haba:2017jgf}, the left-right parity
$\P$ was considered in the limit of decoupled $W_R$, where only the flavor-changing heavy scalar
contributes, making it effectively a particular two Higgs doublet model.  In all cases, the detailed
analysis of the correlations among the different observables shows to be relevant.

In summary, we review and reassess the impact of the $\eps$, $\eps'$ and $n$EDM observables on the
LRSM, paying attention to the theoretical uncertainties, presently dominated by the hadronic matrix
elements, and to the phase patterns and correlations ensuing from either choice of LR symmetry
(generalized $\P$ or $\C$). As far as $\eps'/\eps$ is concerned, we consider two benchmark cases: i)
a scenario in which the SM prediction of $\eps'$ saturates the experimental result, and ii) a
new-physics one where the LRSM contribution is the main source for it.

We conclude that, in the case of $\mathcal{P}$ the standard scenario imposes a lower bound on
the LR scale of $\sim 13\,\TeV$, while a substantial new-physics contribution to $\eps'$ can
arise for $M_{W_R}=7$--$45\,\TeV$, with the $n$EDM at the reach of the new generations of
experiments. The presence of a PQ axion reduces nontrivially the $n$EDM and substantially relaxes these limits.
In the case of $\C$, no lower bound arises in the standard scenario, since the
relevant phases can be set as small as needed. On the other hand, LR contributions can saturate
$\eps'$ for $M_{W_R}$ as large as 100\,\TeV\ according to the configuration of the model parameters.

The study is organized as follows. In the next section we briefly recall the LRSM features which are
relevant for the analysis. In Section~\ref{sec:Kmixing}, \ref{sec:epsilonprime} and
\ref{sec:neutronedm} we review and update the LRSM contributions to the $K^0-\overline{K^0}$
oscillations, $\eps'/\eps$ and $n$EDM respectively, and discuss the status of hadronic matrix elements
calculations.  In section~\ref{sec:numericaldiscussion} we finally show the outcome of our numerical
study.  We report in the appendices the relevant tools, namely loop functions, operator anomalous
dimensions, meson and baryon chiral Lagrangian, and explicit formul\ae\ for the CP-violating phases
in the LRSM.

\section{The Model}
\label{sec:LRSM}

\subsection{The gauge and scalar sectors}

\noindent
The LRSM, with gauge group $SU(2)_L\times SU(2)_R\times U(1)_{B-L}\times SU(3)_c$, contains three
additional gauge bosons related to the $SU(2)_R$ group, $W_R^{\pm}$ and a new neutral vector $Z'$.
Left-handed and Right-handed quarks and leptons are accommodated in the fundamental representations
of $SU(2)_{L,R}$, $Q_{L,R }= \left( u \ d \right)^t_{L,R}$,
$\ell_{L,R} = \left(\nu \ e \right)^t_{L,R}$, with electric charge
$Q = I_{3 L} + I_{3 R} + {B - L \over 2}$, where $I_{3 L,R}$ are the third generators of
$SU(2)_{L,R}$. In analogy with the SM, the RH charged currents induce flavor-violating (FV)
interactions, and furthermore $W_R$ mixes with $W_L$. This provides a RH interaction mediated by the
light mass-eigenstate, mostly the standard gauge boson $W$, namely
\begin{equation}\label{mainL}
\mathcal{L}_{\rm mix-current}= \frac{g}{\sqrt{2}} \zeta \, W^\mu \, \bar{u}_R V_R \gamma_\mu d_R + h.c. \,,
\end{equation}
where $\zeta$ is the gauge boson mixing to be defined shortly, $V_R$ is the right-handed equivalent
of the standard CKM matrix, and $u,d$ span the three quark flavors.

The $SU(2)_R\times U(1)_{B-L}\to U(1)_Y$ spontaneous symmetry breaking is provided by a RH triplet
$\Delta_R (1_L,3_R,2)$
\begin{equation}
\Delta_{R} = \left[ \begin{array}{cc} \Delta^+ /\sqrt{2}& \Delta^{++} \\
\Delta^0 & -\Delta^{+}/\sqrt{2} \end{array} \right]_{R} 
\end{equation}
via the vacuum expectation value (VEV) $v_R$ developed by $\Delta^0$. Then the $W_R$ gauge boson has
mass $M_{W_R}=g v_R$.

The LR mixing $\zeta$ between $W_R$ and $W\simeq W_L$ is
\begin{equation}\label{mix}
\zeta \simeq - r\, e^{i \a}  \sin 2 \b\,,
\end{equation}
with $r={M_{W_L}^2}/{M_{W_R}^2}$ and $ \tan\b\equiv t_\b = v_2/v_1 $. From the direct experimental
limit on the LR scale one obtains $|\zeta| < 4\times 10^{-4}$. Here $v_{1,2}$ are related to the
electroweak breaking, provided by a bidoublet field $\Phi(2_L,2_R,0)$
\begin{equation}\label{HH}
\Phi = \left[\begin{array}{cc}\phi_1^0&\phi_2^+\\\phi_1^-&\phi_2^0\end{array}\right],
\end{equation}
with VEV
$\langle\Phi\rangle = \text{diag}\left\{v_1, e^{i \a} v_2 \right\}$~\cite{Senjanovic:1978ev}. The
standard electroweak VEV is given by $v^2=v_1^2+v_2^2$, with $v\ll v_R$. The standard Higgs boson is
contained in~\eqref{HH} predominantly in the real part of $\phi_1$; the imaginary and complex
components of $\phi_2$ are instead neutral scalars whose masses are proportional to $v_R$. They have
to be heavy enough because they mediate tree-level FCNC, and their presence plays an important role
in the phenomenology of the low-scale LRSM.

For the present study devoted to the CP observables $\eps$, $\eps'/\eps$ and $n$EDM, one of the main
ingredients is Eq.~\eqref{mainL}. It contains sources of CP-violation because of the spontaneous
phase $\a$ inside $\zeta$ and of the complexity of $V_R$. Remarkably, a tree-level contribution to
$\eps'$ is generated by the LR mixing $\zeta$ via an effective four-quark operator (defined below as
$Q_2^{LR}$) obtained after integration of the gauge field. The complete basis of operators, induced
at tree or loop level and through renormalization to low scale, will be listed in
section~\ref{sec:epsilonprime}. A similar treatment is reserved for $n$EDM, in the case of
$\Delta S=0$ transitions: as we shall see, analogous effective operators generate via chiral loops
the dominant contribution to the $n$EDM~\cite{Maiezza:2014ala}. The account and evaluation of the
various sources of $\eps'$ and $n$EDM are the matter of dedicated sections in the following.

\subsection{The choice of LR discrete symmetry}

\noindent
In Eq.~\eqref{mainL} the condition $g_L=g_R=g$ is assumed, being $g_{L,R}$ the gauge coupling of
$SU(2)_{L,R}$. This follows from an additional discrete symmetry in the LRSM relating the left and
right sector. Such a symmetry is not unique: it can be realized either with a generalized parity
$\mathcal{P}$ or a generalized charge conjugation $\mathcal{C}$ which, in addition to exchanging the
weak gauge groups, are defined respectively by~\cite{Maiezza:2010ic}
\begin{equation}\label{PC}
\mathcal{P}: \left\{ \begin{array}{l} Q_L\leftrightarrow Q_R     \\[1ex]  \Phi \to \Phi^\dagger  \end{array}  \right. ,
\qquad
\mathcal{C}: \left\{ \begin{array}{l} Q_L \leftrightarrow (Q_R)^c \\[1ex]  \Phi \to \Phi^T  \end{array}  \right.,
\end{equation}
with analogous transformationa for the lepton doublets.
The action of $\mathcal{P}$ and $\mathcal{C}$ on the Yukawa Lagrangian 
\begin{equation}\label{Yukawa}
{\cal L}_Y  = \overline{Q}_L  \Big(Y\, \Phi \,  + \tilde Y \, \tilde\Phi\Big) Q_R + h.c.
\end{equation}
implies $Y=Y^\dag$ and $Y=Y^T$  respectively, and the same for $\tilde Y$. 
After the quark mass matrices
\begin{eqnarray}
M_u&=& v_1\, Y  + v_2\,{\rm e}^{-i\a}\, \tilde Y \nonumber\\
M_d&=& v_2 \,{\rm e}^{i\a}\, Y + v_1\, \tilde Y
\label{masses}
\end{eqnarray}
are bi-diagonalized given forms of $V_R$ are obtained, according to the properties of $Y$, $\tilde Y$.  
The case of $\mathcal{C}$ is fairly simple~\cite{Maiezza:2010ic}:
\begin{equation}\label{VRC}
V_R=K_u V^* K_d\,,
\end{equation}
with $V$ the standard CKM matrix and $K_{u,d}$ diagonal matrices of free phases
$K_u={\rm diag}\{\e^{i\t_u}, \e^{i\t_c}, \e^{i\t_t}\}$,
$K_d={\rm diag}\{\e^{i\t_d}, \e^{i\t_s}, \e^{i\t_b}\}$, where from now on we adopt $\t_b=0$.

In the case of $\mathcal{P}$, an analytical form for $V_R$ has been recently found, with a
perturbative expansion in the small parameter
$|s_\a\, t_{2\b}|\lesssim 2m_b/m_t\simeq 0.05$~\cite{Senjanovic:2014pva,Senjanovic:2015yea}:
\vspace*{-1ex}
\begin{align}\label{VRP}
V_{R,ij} = & V_{ij}-i s_\a \, t_{2 \b } \Bigg (V_{ij}t_{\b }+ \sum_{k=1}^3 \frac{V_{kj} (V\, m_d \,V^{\dagger })_{ik}}{m_{u\,ii}+m_{u\,kk}} \\ \nonumber
& \qquad \qquad\quad +\frac{V_{ik} (V^{\dagger}\,m_u\,V)_{kj}}{m_{d\,jj}+m_{d\,kk}} \Bigg) + \mathcal{O}(s_\a \, t_{2 \b })^2\,,
\end{align}
where $m_{u,d}$ are the diagonal quark mass matrices.  This expression is not unique, other
solutions are found by replacing $m_{ii}\to s_i m_{ii}$ and
\beq
 V\rightarrow {\rm diag}\{s_u,s_c,s_t\} \,V\, {\rm diag}\{s_d,s_s,s_b\}\,, 
 \label{phasesigns}
\eeq
where $s_i$ are arbitrary signs (and from now on we adopt $s_b=1$).  In Appendix~\ref{app:phases},
explicit expressions for the relevant phase combinations are given for generic $s_i$.

The argument of the determinant of the fermion mass matrices can also be
computed~\cite{Senjanovic:2014pva,Senjanovic:2015yea,Maiezza:2014ala}, namely
\beq\label{eq:thetabar}
\bar\theta\simeq \frac12 s_\a \, t_{2 \b } \, \Re\,{\rm tr}\left(m_u^{-1}V m_d V^\dag -m_d^{-1}V^\dag m_u V \right)  
\eeq
at the first order in $s_\a\, t_{2 \b }$.

\section{$K^0-\bar K^0$ mixing}
\label{sec:Kmixing}

\noindent
A particularly constraining process for the LRSM is the neutral kaon mixing, effectively induced
through the chirally enhanced operator
\begin{equation}
  \sand{K}{^0}{\bar{s} L d\, \bar{s} R d} =  \frac12 f_K^2 m_K {\mathcal B}^K_4\left[\frac{m_K^2}{(m_s+m_d)^2} + \frac{1}{6}\right]\,,
\end{equation}
where $f_{K}$ and $m_{K}$ are the decay constant and the mass of the meson $K$ respectively, and
$L,R=(1\mp\gamma_5)/2$. This operator is generated via the LR box diagrams and even at the
tree-level through the exchange of a flavor-changing (FC)
scalar~\cite{Senjanovic:1978ev,Senjanovic:1979cta}. The inclusion of the one-loop renormalization of
the tree-level diagrams, necessary for a gauge invariant result~\cite{Basecq:1985cr}, has
phenomenologically relevant implications~\cite{Bertolini:2014sua}. The bag factor ${\mathcal B}^K_4$
has been computed on the lattice by various groups with some $20\%$
discrepancies~\cite{Aoki:2019cca}. For the present study, we follow the discussion and numerical
analysis of Ref.~\cite{Bertolini:2014sua}, to which we refer the reader for the details.

The CP violating parameter $\eps$, which gauges the indirect CP violation in the mixing, is
particularly important in constraining the external CP-phases introduced in the previous
section~\cite{Maiezza:2010ic,Bertolini:2014sua}. The bounds may be inferred by using a convenient
parametrization of new physics in $\eps$, namely
\begin{equation}
h_{\eps} \equiv \frac{\Im\sand{K}{^0}{\H_{LR}}}{\Im\sand{K}{^0}{\H_{LL}}}\, ,
\end{equation}
which in the case of $\C$ turns into
\begin{equation}
h^\C_{\eps}\simeq{\rm Im}\!\left[e^{i(\theta_d-\theta_s)}\big( A_{cc}+\!A_{ct}\cos(\theta_{c}-\theta_{t}+\phi)\big)\right],
\end{equation}
while in the case of $\P$ becomes
\begin{equation}
h^\P_{\eps} \propto {\rm Im}\!\left[e^{i(\theta_d-\theta_s)}\left[ A_{cc}+\!A_{ct}e^{i \phi}\cos(\theta_{c}-\theta_{t}) \right]\right],
\end{equation}
with $\phi=\arg(V_{Ltd})\simeq -22^\circ$. $A_{cc,ct}$ correspond to the contributions of
charm-charm and charm-top quark in the effective Lagrangian. They are real numbers which scale circa
as $M_{W_R}^{-2}$, as we consider the contribution of the FC scalar to be at most comparable to the
one of $M_{W_R}$, corresponding to $M_H\simeq 6M_{W_R}$ within the perturbative
regime~\cite{Bertolini:2014sua}. For a wide range of $M_{W_R}$ one has $A_{ct}/A_{cc}\simeq 0.45$.

Conservatively we allow the amount of new physics in $\eps$ to be at most
10\%~\cite{Buras:2013ooa}.  This translates into a sharp constraint on $\theta_d-\theta_s$~\cite{Bertolini:2014sua}, which
in the case of $\C$ reads 
\begin{align}
  \label{eq:epsC}
  &|\sin(\theta_s-\theta_d)|_{s_cs_t=-1} < \left(\frac{M_{W_R}}{104\,\TeV}\right)^2 \nonumber \\
&|\sin(\theta_s-\theta_d)|_{s_cs_t=1} < \left(\frac{M_{W_R}}{71\,\TeV}\right)^2\,,
\end{align}
while for $\P$ one has
\begin{align}
  \label{eq:epsP}
  &|\sin(\theta_s-\theta_d+0.16)|_{s_cs_t=-1} < \left(\frac{M_{W_R}}{104\,\TeV}\right)^2  \nonumber \\
&|\sin(\theta_s-\theta_d-0.16)|_{s_cs_t=1} < \left(\frac{M_{W_R}}{71\,\TeV}\right)^2\,.
\end{align}

\section{Direct CP violation in $K^0\to\pi\pi$}
\label{sec:epsilonprime}

\subsection{Effective interactions}

\noindent
Mesonic and hadronic processes that involve weak interactions are best described in terms of the
operator product expansion, which factorizes short- and long-distance effects.
 For $\Delta S=1$
flavor changing transitions the effective Lagrangian can be written in the form
\beq
L_{\Delta S=1} =-\frac{G_F}{\sqrt{2}}\, \sum_i C_iQ_i+h.c.\,,
\label{DS1Lag}
\eeq
where $Q_i$ are the relevant operators and $C_i$ the corresponding Wilson coefficients ($G_F$ is the Fermi
constant).  

In the Standard Model the $\Delta S=1$ Lagrangian involves tree-level operators as well as QED and
QCD induced loop diagrams. When both left and right chirality interactions are present, the standard
set of operators is enlarged to include, at the scale of 1 GeV, 28 operators~\cite{Bertolini:2012pu}
\begin{equation}
\begin{array}{ll}
Q_1^{LL} = (\bar{s}_\a u_\b)_L (\bar{u}_\b d_\a)_L & Q_1^{RR} = (\bar{s}_\a u_\b)_R (\bar{u}_\b d_\a)_R \\[0.2cm]
Q_2^{LL} = (\bar{s} u)_L (\bar{u} d)_L & Q_2^{RR} = (\bar{s} u)_R (\bar{u} d)_R \\[0.2cm]
Q_3      = (\bar{s} d)_L (\bar{q} q)_L &Q'_3      = (\bar{s} d)_R (\bar{q} q)_R \\[0.2cm]
Q_4      = (\bar{s}_\a d_\b)_L (\bar{q}_\b q_\a)_L&Q'_4      = (\bar{s}_\a d_\b)_R (\bar{q}_\b q_\a)_R \\[0.2cm]
Q_9      = \frac{3}{2}(\bar{s} d)_L e_q(\bar{q} q)_L&Q'_9      = \frac{3}{2}(\bar{s} d)_R e_q(\bar{q} q)_R \\[0.2cm]
Q_{10}      = \frac{3}{2}(\bar{s}_\a d_\b)_L e_q(\bar{q}_\b q_\a)_L&Q'_{10}      = \frac{3}{2}(\bar{s}_\a d_\b)_R e_q(\bar{q}_\b q_\a)_R 
\end{array}\!\!\!\!\!\!
\label{eq:DS1operators1}
\end{equation}

\leavevmode
\vspace*{-2.3em}

\begin{equation}
\begin{array}{ll}
Q_1^{RL} = (\bar{s}_\a u_\b)_R (\bar{u}_\b d_\a)_L & Q_1^{LR} = (\bar{s}_\a u_\b)_L (\bar{u}_\b d_\a)_R \\[0.2cm]
Q_2^{RL} = (\bar{s} u)_R (\bar{u} d)_L &Q_2^{LR} = (\bar{s} u)_L (\bar{u} d)_R \\[0.2cm]
Q_5      = (\bar{s} d)_L (\bar{q} q)_R&Q'_5      = (\bar{s} d)_R (\bar{q} q)_L \\[0.2cm]
Q_6      = (\bar{s}_\a d_\b)_L (\bar{q}_\b q_\a)_R&Q'_6      = (\bar{s}_\a d_\b)_R (\bar{q}_\b q_\a)_L \\[0.2cm]
Q_7      = \frac{3}{2}(\bar{s} d)_L e_q(\bar{q} q)_R&Q'_7      = \frac{3}{2}(\bar{s} d)_R e_q(\bar{q} q)_L \\[0.2cm]
Q_{8}      = \frac{3}{2}(\bar{s}_\a d_\b)_L e_q(\bar{q}_\b q_\a)_R\ &Q'_{8}      = \frac{3}{2}(\bar{s}_\a d_\b)_R e_q(\bar{q}_\b q_\a)_L\  
\end{array}\!\!\!\!\!\!
\label{eq:DS1operators2}
\end{equation}

\leavevmode
\vspace*{-2.3em}

\begin{equation}
\begin{array}{ll}
Q_{g}^{L}  = \frac{g_sm_s}{16\pi^2}\bar{s}\sigma_{\mu\nu}t^a G_a^{\mu\nu}L d\ \ & Q_{g}^{R}  = \frac{g_sm_s}{8\pi^2}\bar{s}\sigma_{\mu\nu}t^a G_a^{\mu\nu}R d \\[0.2cm]
Q_{\g}^{L}  = \frac{em_s}{16\pi^2}\bar{s}\sigma_{\mu\nu} F_a^{\mu\nu}L d \ \ & Q_{\g}^{R}  = \frac{em_s}{8\pi^2}\bar{s}\sigma_{\mu\nu} F_a^{\mu\nu}R d\, ,
\end{array}\!\!\!\!
\label{eq:DS1operators3}
\end{equation}
with $(\bar{q}q)_{L,R}\equiv \bar q \gamma_\mu (L,R) q$, $L,R\equiv 1\mp\gamma_5$, and implicit
summation on $q=u,d,s$. $Q_{1,2}^{LL}$ are the SM operators usually denoted as $Q_{1,2}$. The dipole
operators $Q_{g,\gamma}$ are normalized with $m_s$, for an easy comparison with existing
calculations and anomalous dimensions. It is known that some of the operators above are
characterized by enhancements due to their chiral structure, either in the running of the short
distance coefficient or in the matrix element. In particular the Wilson coefficients of the QCD
dipole operators $Q_g^{L,R}$ receive a large enhancement from the mixing with the current-current
operators.

At the leading order the operators generated by the SM and the LR short distance physics are:
$Q_2^{AB}$, $Q_4$, $Q'_4$, $Q_6$, $Q'_6$, $Q_7$, $Q'_7$, $Q_9$, $Q'_9$, $Q^{A}_{g}$, $Q^{A}_{\g}$,
with $A,B=L,R$.
Their Wilson coefficients are for current-current operators
\beq
\begin{split}
C_2^{LL} = \lambda_u^{LL}\,,\qquad&
C_2^{LR} = \zeta^* \lambda_u^{LR}\,,\\[.7ex]
C_2^{RR} = r\ \lambda_u^{RR}\,,\;\quad&
C_2^{RL} = \zeta \,\lambda_u^{RL}\,;
\end{split}
\label{eq:C2AB}
\eeq
for the penguins
\beq
\begin{split}
&C_4=C_6  = \frac{\a_s}{4\pi} \Sigma_i \lambda_i^{LL} F_1^{LL}(x_i)\\[.7ex]
&C'_4=C'_6  = \frac{\a_s}{4\pi} r \, \Sigma_i \lambda_i^{RR} F_1^{RR}(r x_i)\\[.7ex]
&C_7=C_9  = \frac{\a e_u}{4\pi} \Sigma_i \lambda_i^{LL} E_1^{LL}(x_i)\\[.7ex]
&C'_7=C'_9  = \frac{\a e_u}{4\pi} r\, \Sigma_i \lambda_i^{RR} E_1^{RR}(r x_i)\,;
\end{split}
\label{eq:C4C9}
\eeq
and for the dipoles
{\small
\arraycolsep=.2em
\beq
\begin{split}
\!\!\!\!&m_sC_{g}^{L}\! = \Sigma_i\Big[ m_s \lambda_i^{LL} F^{LL}_2 +\!\zeta  m_i \lambda_i^{RL} F^{LR}_2+\!m_d r \lambda_i^{RR} F^{RR}_2\Big]\!\! \\[1ex]
\!\!\!\!&m_sC_{g}^{R} \!= \Sigma_i\Big[ m_d \lambda_i^{LL} F^{LL}_2 + \!\zeta^*  m_i \lambda_i^{LR} F^{LR}_2+\!m_s r \lambda_i^{RR} F^{RR}_2\Big]\!\!\! \\[1ex]
\!\!\!\!&m_sC_{\g}^{L} \!=\Sigma_i\Big[ m_s\lambda_i^{LL} E^{LL}_2 +\!\zeta  m_i \lambda_i^{RL} E^{LR}_2+\!m_d r \lambda_i^{RR} E^{RR}_2\Big]\!\!\! \\[1ex]
\!\!\!\!&m_sC_{\g}^{R} \!= \Sigma_i\Big[ m_d \lambda_i^{LL} E^{LL}_2 +\!\zeta^*  m_i \lambda_i^{LR} E^{LR}_2+\!m_s r \lambda_i^{RR} E^{RR}_2\Big]\!\rlap{\,\,.}
\end{split}
\label{eq:Cg12}
\eeq
}%
In the above, $e_u=2/3$ is the $u$-quark charge, $x_i=m_i^2/m_{W_L}^2$ with $i=u,c,t$, and
$F^{AB}_{1,2}$ and $E^{AB}_{1,2}$ are the loop functions, given in appendix~\ref{app:loop}.
The parameters $\zeta$ and $r$ are defined in \eq{mix}.
Finally
$\lambda_i^{AB}=V_{is}^{* A}V_{id}^{B}$, where $V_L$ and $V_R$ are the Cabibbo-Kobayashi-Maskawa
(CKM) matrix and its right-handed analogue (\eqs{VRC}{VRP}).  

The different terms of the coefficients in \eqs{eq:C2AB}{eq:Cg12} are generated at the decoupling of
the relevant heavy thresholds, and thus at different scales, namely $M_{W_L}$ or $m_t$ for the
current-current operators and top-dominated loops, $m_c$ for the charm dominated loops, and
$m_{W_R}$ for the RR current-current ones.

The direct CP violation in $K^0\to\pi\pi$ decays is parametrised as
\begin{equation}\label{epsilonprimeoverepsilon}
  \Re\frac{\eps'}{\eps}\simeq \frac{\omega}{\sqrt{2}\ |\eps|}\,\left(\frac{{\rm Im} A_2}{{\rm Re} A_2}-\frac{{\rm Im} A_{0}}{{\rm Re} A_{0}}\right),
\end{equation}
where $\omega= \Re A_{2}/\Re A_{0}\simeq 1/22.2$. The isospin amplitudes $A_{I}$ ($I=0,2$) are
defined from the $\Delta S=1$ effective Hamiltonian as
$\langle (2\pi)_{I}|(-i)H_{\Delta S=1}|K^{0}\rangle =A_{I}e^{i\delta_{I}}$, where $\delta_{I}$ are
the strong phases of $\pi \pi$ scattering. The phase of $\eps'$,
$\pi/2+\delta_{2}-\delta_{0}=42.5^0\pm0.9^0$, cancels to a very good approximation the phase of
$\eps$.

While the imaginary part of the amplitudes are calculated within the model, the real parts are set
at their experimental values: $\Re A_0=3.33\times 10^{-7}\,\GeV$ and
$\Re A_2=1.49\times 10^{-8}\,\GeV$, as well as the indirect CP violation parameter
$|\eps|=(2.228 \pm 0.011) \times 10^{-3} $. Because of the large uncertainty associated to the new
physics contribution to $\eps'$ we neglect in \eq{epsilonprimeoverepsilon} a $O(10\%)$ isospin
breaking correction (for a recent recap on isospin violation in the SM amplitudes see
Ref.~\cite{Gisbert:2017vvj}). As a matter of fact, the major source of uncertainty resides in the
evaluation of the hadronic matrix elements, that we are going to discuss next.

For the following discussion and numerical analysis it is convenient to introduce the parameter
\begin{equation}
h_{\eps'} =  \frac{\eps'_{LR}}{\eps'_{\rm exp}}\, ,
\label{hepsprime}
\end{equation}
where $\eps'_{LR}$ represents the additional LRSM contribution to $\eps'$, and is normalized to
the present experimental central value, $|\eps'_{\rm exp}|=3.7\times 10^{-6}$.

\subsection{Matrix elements}
\label{matrixelements}

\noindent
In this section we address the evaluation of the $K^0\to\pi\pi$ matrix elements of the left-right
current-current operators $Q^{LR}_{1,2}$.  We define
$\vev{Q^{LR}_i}_{0,2}\equiv \bra{(\pi\pi)_{I=0,2}} Q^{LR}_i \ket{K^0}$. A naive estimate is provided
by the simple factorization of the matrix elements in terms of currents and densities and vacuum
insertion known as the Vacuum Saturation Approximation (VSA)~\cite{Gaillard:1974hs}.  In spite of
the expected large non factorizable corrections the VSA has been conveniently used in the past as a
reference benchmark. The calculation of the current-current operators in the left right framework
via the VSA prescription is found in \cite{Ecker:1985vv}. In terms of the 
$Q^{LR}_{1,2}$ operators defined above one has
\bea
\langle Q_1^{LR}\rangle_0 &&=-\langle Q_1^{RL}\rangle_0= -\frac{1}{3} \sqrt{\frac{2}{3}} (X+9 Y+3 Z)\nonumber\\[0ex]
\langle Q_1^{LR}\rangle_2 &&=-\langle Q_1^{RL}\rangle_2= -\frac{1}{3} \sqrt{\frac{1}{3}} (X-6 Z)\nonumber\\[0ex]
\langle Q_2^{LR}\rangle_0 &&=-\langle Q_2^{RL}\rangle_0= -\frac{1}{3} \sqrt{\frac{2}{3}} (3 X+3 Y+Z)\nonumber\\[0ex]
\langle Q_2^{LR}\rangle_2 &&=-\langle Q_2^{RL}\rangle_2= -\frac{1}{3} \sqrt{\frac{1}{3}} (3 X-2 Z)\, ,
\label{eq:QLR02}
\eea
with
\begin{align}
X & \equiv i \langle \pi^{+}|\bar{u}\gamma_{\mu}\gamma_{5}d|0\rangle\langle\pi^{-}|\bar{s}\gamma^{\mu}u|{K}^{0}\rangle\simeq\sqrt{2}f_{\pi}(m_{K}^{2}-m_{\pi}^{2}) \notag \\[.7ex]
Y & \equiv i \langle\pi^{+}\pi^{-}|\bar{u}u|0\rangle\langle 0|\bar{s}\gamma_{5}d|{K}^{0}\rangle\simeq\sqrt{2}f_{K} m_{K}^{4}/(m_{s}+m_{d})^2 \notag \\[.7ex]
Z & \equiv i \langle \pi^{+}|\bar{u}\gamma_{5}d|0\rangle\langle\pi^{-}|\bar{s}u|{K}^{0}\rangle\simeq\sqrt{2}f_{\pi} m_{K}^{4}/(m_{s}+m_{d})^2\,.
\label{eq:XYZ}
\end{align}
At variance with \cite{Ecker:1985vv} in \eq{eq:XYZ} a factor $i$ is conventionally factored out.
When allowed, charged pions are replaced by neutral pions with a factor of 2 accounting for their
exchange, so that for instance
$iZ = 2 \langle \pi^{0}|\bar{u}\gamma_{5}u|0\rangle\langle\pi^{0}|\bar{s}d|{K}^{0}\rangle$. The term
$Y$ contributes equally to charged and neutral pions and accordingly it is absent in the isospin 2
projection of the amplitudes
\bea
 A_0 &= \frac{1}{\sqrt{6}}(2A_\pm + A_{00})\,,\nonumber\\[0ex]
 A_2 &= \frac{1}{\sqrt{3}}(A_\pm - A_{00})\,.
 \label{eq:A0A2}
\eea
In \eq{eq:QLR02} non leading terms in $1/N$ due to color Fierz are kept according to the VSA
prescription.  For $Q_2^{LR}$ the $1/N$ terms $Y$ and $Z$ are chirally enhanced and
dominate the amplitude, that turns out to be approximately $1/3$ of the corresponding one of
$Q_1^{LR}$. The VSA values in \Table{matrixel} are given at the scale of 1~GeV with
$(m_s+m_d)(1\ \rm GeV)\approx 132\,$MeV.

\begin{table}[t]
  \centering
  \renewcommand{\arraystretch}{1.5}
  \begin{tabular}{@{} lrrrr @{}}
 \hline
  \hspace*{8ex}   & \qquad VSA& \qquad  $\chi$QM& \qquad  DQCD \\
 \hline
    $\vev{Q^{LR}_1}_0$ & $-1.8$   & $-3.6$  & $-1.1$  \\
    $\vev{Q^{LR}_1}_2$ & $0.53$  & $0.33$ & $0.40$  \\
    $\vev{Q^{LR}_2}_0$ & $-0.62$ & $-1.2$  & $-0.059$  \\
    $\vev{Q^{LR}_2}_2$ & $0.16$  & $0.092$ & $-0.005$  \\ [1ex]
 \hline
  \end{tabular}
  \caption{Comparison of $K^0\to\pi\pi$ matrix elements of the left-right current-current operators
    $Q^{LR}_{1,2}$ in different approaches.  The values are given at the scale of 1~GeV in units of
    $\rm GeV^3$ for central values of the relevant input parameters.\label{matrixel} }
\end{table}

In the second column we report the results of the computation of the $K^0\to\pi\pi$ matrix elements
of $Q^{LR}_{1,2}$ within the Chiral Quark Model ($\chi$QM)
approach~\cite{Bertolini:2012pu,Bertolini:2013noa}. In this modeling of low energy QCD the meson
octet chiral Lagrangian is complemented with an effective quark-meson
interaction~\cite{Manohar:1983md,Cohen:1984vv}, which provides a connection between the QCD degrees
of freedom and the lightest hadronic states. Meson interactions are then obtained by integration of
the constituent quarks and the chiral Lagrangian coefficients are determined in terms of three
non-perturbative parameters, namely the constituent quark mass and the quark and gluon condensates.

In the nineties an extensive program was carried out in order to calculate all $\Delta S=1$ matrix
elements relevant to the $\Delta I=1/2$ rule and direct CP violation in $K^0\to\pi\pi$ decays based
on the $\chi$QM at the NLO in the chiral expansion~\cite{Bertolini:1998vd}. By adopting a
phenomenological approach it was shown that a fit of the $\Delta I=1/2$ rule could be obtained for
expected values of the three non-perturbative parameters of the
model~\cite{Bertolini:1997ir}.\footnote{Model dependent non-factorizable $1/N$ corrections
  proportional to the gluon condensate were shown to play a crucial role in depleting the isospin 2
  amplitude while contributing to the enhancement of the isospin 0 amplitude.} In turn, this allowed
a coherent calculation of the matrix elements for the whole dimension six $\Delta S=1$ SM
Lagrangian\cite{Antonelli:1995nv}, including the dimension five chromomagnetic dipole
operator~\cite{Bertolini:1993rc,Bertolini:1994qk}.\footnote{Very recent
  lattice~\cite{Constantinou:2017sgv} and QCD model~\cite{Buras:2018evv} calculations of the gluon
  dipole operator give a $K^0\to\pi\pi$ matrix element smaller by about a factor of
  two~\cite{Bertolini:2012pu}.}

The relevance of non-factorizable 1/N model and chiral corrections in lifting the cancellation
between the gluon and electromagnetic penguin was exposed, leading in 1998 to the prediction
$\eps'/\eps=17\, ^{+17}_{-10}\times 10^{-4}$~\cite{Bertolini:1995tp,Bertolini:1997nf} shortly
afterwards confirmed by the precise experimental findings of KTEV~\cite{AlaviHarati:1999xp} and
NA48~\cite{Fanti:1999nm} collaborations
\beq
\eps'/\eps = 16.6 \pm 2.3 \times 10^{-4}\,.
\label{epspexp}
\eeq

Subsequent attempts to resum the final state interactions via dispersion
relations~\cite{Pallante:1999qf,Pallante:2000hk,Pallante:2000pz,Buchler:2001nm,Buchler:2001np} lead
to a confirmation of the enhancement of the $I=0$ amplitudes and of the agreement between SM and
data.

In recent years the phenomenological $\chi$QM framework has been applied to the calculation of the
matrix elements of relevant operators in the left-right model~\cite{Bertolini:2013noa}. In
\Table{matrixel} the matrix elements of $Q^{LR}_{1,2}$ obtained in the model at the scale of 0.8~GeV
are evolved to 1~GeV. As remarked in~\cite{Bertolini:2013noa}, attention must be paid in subtracting
an unphysical contribution to the $K^0\to\pi\pi$ amplitudes generated by the presence of a
$K^0\to\,$vacuum transition (tadpole)~\cite{Feinberg:1959ui} induced by the LR current-current
operators. We see again in the $\chi$QM calculation an enhancement of the $\Delta I=1/2$ amplitudes
compared to the $\Delta I=3/2$ ones. This pattern is led by the one-loop chiral loop contributions,
which include the final state rescattering.~\footnote{The impact of final-state interactions in
  $\eps'/\eps$ has been recently questioned in~\cite{Buras:2016fys}}

In the third column of \Table{matrixel} we report the results obtained with the Dual QCD (DQCD)
approach~\cite{Aebischer:2018rrz} (for a recent summary and references see
Ref.~\cite{Buras:2018ozh}). A rescaling factor of $\sqrt{3/2}$ has been applied in order to
normalize the DQCD results to the amplitudes defined in \eq{eq:A0A2}. The DQCD calculation of the
hadronic matrix elements is based on a truncated chiral Lagrangian and leading $N$
factorization~\cite{Bardeen:1986vp,Bardeen:1986uz,Bardeen:1986vz}. The matrix elements of the four
quark operators are calculated in the large $N$ limit at zero momentum by factorizing the four quark
operators in terms of color singlet currents or densities via their chiral representations. The
meson operators undergo an evolution quadratic in the cutoff scale up to
$\Lambda\simeq 0.7$~GeV~\cite{Fatelo:1994qh}. They are then matched with the short-distance Wilson
coefficients at the 1~GeV scale.

This approach has shown to be successful in the past in predicting the size of the bag parameter
$B_K$ in $\bar K^0-K^0$ mixing~\cite{Bardeen:1987vg,Buras:2014maa} in agreement with lattice
calculations.~\cite{Garron:2016mva,Boyle:2017skn} Recently, by supporting (and providing a model
rationale for~\cite{Buras:2018lgu}) the RBC-UKQCD lattice results~\cite{Blum:2015ywa,Bai:2015nea}
for the $\Delta I=1/2$ rule and the direct CP violation in $K^0\to\pi\pi$ decays, leading to
\beq
\eps'/\eps = 1.38 \pm 6.90 \times 10^{-4}\,.
\label{epspLattice}
\eeq

In spite of the enormous progress made in the past decade, the present lattice calculations of the
$K\to \pi\pi$ matrix elements still fail in reproducing the strong rescattering phase $\delta_0$ (by
about 3$\sigma$) and do not include isospin breaking.~\footnote{The role of isospin breaking
  contributions in the lattice calculations has been further scrutinized
  in~\cite{Aebischer:2019mtr}.}
It is fair to say that, given the delicate cancellation between the QCD and QED penguin operators
that leads to the present SM estimate, we should await for a comprehensive and precise lattice
description of the $K\to \pi\pi$ decays before claiming the need of new physics explanations.

On the other hand, a detailed reevaluation of $\eps'/\eps$ within the chiral Lagrangian framework,
including isospin breaking, leads to~\cite{Gisbert:2017vvj}
\beq
\eps'/\eps = 15 \pm 7 \times 10^{-4}
\label{epspLchi}
\eeq
in agreement, albeit with a large error, with the data.\footnote{A very recent update based on a
  detailed reassessment of the isospin breaking effects leads to
  $\eps'/\eps = 14 \pm 5 \times 10^{-4}$~\cite{Cirigliano:2019cpi}.}

In \Table{matrixel} the values of the relevant matrix elements are reported at the scale of 1\,GeV
for our operator basis. It is apparent in the comparison the reduced size of $\vev{Q^{LR}_1}_0$ and
the minuscule size of the $Q^{LR}_2$ matrix elements. These results stem from the large $N$
factorization of the current-current operator (corresponding to the subleading term $X$ in
\eq{eq:XYZ}). The meson evolution of the $Q^{LR}_2$ operator mixes it with the chirally enhanced
$Q^{LR}_1$ but with a renormalization suppression factor of $\Lambda^2/(4\pi f_\pi)^2$, so that
$1/N$ chirally enhanced terms turn out to be not effective. This is a distinctive feature of the
DQCD approach.

The authors of Ref.~\cite{Haba:2018byj} invoke isospin symmetry to connect the $Q^{LR}_{1.2}$ matrix
elements to the analogues of the SM gluon and electromagnetic penguins $Q_{5,6,7,8}$, for which
lattice calculation are available at a scale of 3~GeV. No numerical details are given
in~\cite{Haba:2018byj}. We find that the $Q^{LR}$ matrix elements so derived follow quite nearly the
pattern and size of the VSA results.

Given the spread and pattern of values in \Table{matrixel} we conservatively use in our analysis the results of VSA as a reference benchmark, while 
including a conservative theoretical uncertainty of a factor of two.

\section{Neutron EDM}
\label{sec:neutronedm}

\subsection{Strong CP in LRSM}

\noindent
While the SM provides a natural answer to the smallness of $\bar\theta$, the latter being
perturbatively protected~\cite{Ellis:1978hq}, more general approaches have been proposed, which are
relevant for new physics extensions.  The Peccei-Quinn (PQ) axion
models~\cite{Peccei:1977hh,Weinberg:1977ma,Wilczek:1977pj} provide an elegant way to address
dynamically the problem. On the other hand other solutions involving the UV completion are possible,
as the restoration of a mirror symmetry in the fermion sector~\cite{Barr:1991qx} or other
extensions~\cite{Kuchimanchi:2010xs,Kuchimanchi:2018ebf}.  For a grand unified embedding see
\cite{Mimura:2019yfi}.

Within the LRSM such a solution is provided by the scenario in which the $\P$ symmetry is exact at
high scale and then spontaneously broken~\cite{Mohapatra:1978fy}.
The symmetry sets to zero the
topological term $\theta$, so that $\bar{\theta}$ is computable after spontaneous breaking, see
\eq{eq:thetabar}.  Since this is by far the dominant contribution to $d_n$, the constraint on
$\bar{\theta}$ translates into a very stringent limit on the combination $s_\a t_{2\b}$ and thus
into the effective vanishing of all phases $\theta_i$, which are directly driven by it.  In such a
situation the $\eps$ constraint in~\eq{eq:epsP} implies a lower bound on the LR scale,
$M_{W_R}\gtrsim 28\,\TeV$~\cite{Maiezza:2014ala}, as derived in the limit
$\t_s-\t_d\to0$~\cite{Bertolini:2014sua}. This conclusion is avoided if $\bar\t$ is canceled
 by a different mechanism, like the PQ one.
In the case of LRSM with $\C$, both $\arg \det M$ and $\theta_{QCD}$ are free parameters, and if one
does not want to exploit this freedom as a fine-tuning, the $\bar{\theta}$ issue has again to be
addressed by assuming some underlying mechanism, as mentioned above.

After the PQ setup removes $\bar{\t}$, still the presence of P- and CP-violating LR effective
operators generates various sources of the $n$EDM~\cite{An:2009zh}, in both the $\C$ and $\P$
cases. We review and compare their impact in the following sections.  It turns out that the most
relevant contribution to $d_n$ is due to meson loops after the shift of the meson fields in the
$U(3)$ chiral Lagrangian, induced by the CP-violating four-quark operators~(see
Appendix~\ref{app:Lchiral}).  As we will see, the proper calculation of the meson-baryon couplings
in the $U(3)$ chiral Lagrangian, shows in the PQ case an exact cancelation that suppresses the
predicted $n$EDM. This result is due to the remnant $\bar\t$ induced by the relevant LR quark
operators. Such a feature was missed in Ref.~\cite{Haba:2018byj}, where a different dependence of
the meson-baryon couplings on the mesons VEVs is obtained.

In the following sections we briefly review the contributions to the neutron EDM, arising in the
LRSM, from short- and long-distance sources.

\subsection{Effective operators}

\noindent
The effective CP odd Lagrangian relevant for the $n$EDM can be written as~\cite{Khatsimovsky:1987fr,Xu:2009nt}
\bea
{\cal L}_{EDM}&=&-\frac{G_F}{\sqrt{2}}\left( \sum_{q\neq q', i=1}^2  {\cal C}_{i}^{qq'} {\cal O}_{i}^{qq'}
+ \frac{1}{2} \sum_{q\neq q', i=3}^4  {\cal C}_{i}^{qq'} {\cal O}_{i}^{qq'}
\right. \nn\\
&&{}\qquad\quad + \left. \sum_{q, i=1}^4  {\cal C}_{i}^{q} {\cal O}_{i}^{q}
+{\cal C}_5 {\cal O}_5\right),
\label{eq:Ledm}
\eea
where $q=u,d,s$ and the effective operators are given by
\bea
&&{\cal O}_{1}^{q'q}=\bar{q}'q' \ \bar{q}i\gamma_5q,
\ \ \ {\cal O}_{2}^{q'q}=\bar{q}'_\a q'_\b \ \bar{q}_\b i\gamma_5q_\a,
\label{eq:O1}
\\
&&{\cal O}_{3}^{q'q}=\bar{q}'\sigma^{\mu\nu}q' \ \bar{q}\sigma_{\mu\nu}i\gamma_5q,
\\
&&{\cal O}_{4}^{q'q}=\bar{q}'_\a \sigma^{\mu\nu}q'_\b \ \bar{q}_\b\sigma_{\mu\nu} i\gamma_5q_\a,
\\
&&{\cal O}_{1}^{q}=\bar{q}q \ \bar{q}i\gamma_5q, \ \ \
{\cal O}_{2}^{q}=\bar{q}\sigma_{\mu\nu}q \ \bar{q}\sigma^{\mu\nu}i\gamma_5q,
\\
&&{\cal O}_{3}^{q}=-\frac{e}{16\pi^2}e_q\,m_q\bar{q}\sigma_{\mu\nu}i\gamma_5q\,F^{\mu\nu},
\\
&&{\cal O}_{4}^{q}=-\frac{g_s}{16\pi^2}\,m_q\bar{q}\sigma_{\mu\nu}i\gamma_5T^aq\,G^{a\,\mu\nu},\label{eq:dipole}
\\
&&{\cal O}_5=-\frac{1}{3} \frac{g_s}{16\pi^2} f^{abc}
G^a_{\mu\sigma}G^{b,\sigma}_{\nu}\widetilde G^{c,\mu\nu}\,.
\label{eq:Weinberg}
\eea
The tensor operators ${\cal O}_{3,4}^{q'q}$ are symmetric in $q'\, q$
($i\gamma_5\sigma^{\mu\nu}\propto \eps^{\mu\nu\a\b}\sigma_{\a\b}$), hence the factor $1/2$ in
\eq{eq:Ledm}. The ${\cal O}_{1,2}^{q'q}$ and dipole operators ${\cal O}_{3,4}^{q}$ are obtained from
the $\Delta S=1$ Lagrangian \eqs{DS1Lag}{eq:DS1operators3} by replacing $s\to d=q$. Accordingly, the
Wilson coefficients $C_{1,2}^{q'q}$ at the weak scale are related to the $\Delta S=1$ ones by
\beq
C_{1,2}^{ud}=-C_{1,2}^{du}=4\, \Im C_{1,2}^{RL}\, ,
\label{C12ud-LR}
\eeq
while
\bea
m_{u,d}\, e_{u,d}\, C_{3}^{u,d}&=& 2 m_s\, \Im (C_{\gamma}^{R,L})\, , \\[1ex]
m_{u,d}\, C_{4}^{u,d}&=& 2 m_s\,  \Im (C_{g}^{R,L})\, ,
\label{C34-LR}
\eea
where one should replace  $\lambda_{i= q'}^{AB}$ with $\lambda_{q'q}^{AB}=V_{q'q}^{* A}V_{q'q}^{B}$, thus  selecting in \eq{eq:Cg12} only the mixed LR terms.

The Wilson coefficient of the three-gluon operator ${\cal O}_5$ is suppressed by $\a_s/{4\pi}$ and
its contribution to the $n$EDM is negligible for light quarks. At the integration scale of each heavy
quark it is given by~\cite{Hisano:2012cc,Braaten:1990gq,Chang:1991ry,Chang:1992vs}
\beq
C_{5}(m_q)= \frac{\a_s(m_q)}{8\pi} C_{4}^{q}(m_q)\, ,
\label{C5-LR}
\eeq
with $q=b$ giving the dominant contribution, proportional to $m_t/m_b$ (see \eq{eq:Cg12}).

By inspection of \eqs{eq:C2AB}{eq:Cg12}, the leading operators induced by gauge boson exchange, which are
sensitive to the new CP phases through the LR mixing $\zeta$, are those obtained from $Q_{2}^{LR,RL}$ and
$Q_{g,\gamma}^{L,R}$.  The ${\cal O}_{1,2}^{q}$ operators are induced by neutral scalar exchange
with CP-violating couplings ($Z$ boson exchange does not induce $CP$ violating transitions).  On the
other hand, the contributions of the heavy doublet Higgs, that we assume decoupled at a scale higher
than the right-handed gauge bosons, are always suppressed by the light quarks Yukawa couplings, and
are henceforth neglected.
Analogously, the operators ${\cal O}_{3,4}^{q'q}$ are not generated at the tree level in the model,
but obtained via gluonic corrections of ${\cal O}_{1,2}^{q'q}$. Since ${\cal O}_{3,4}^{q'q}$ are
flavor symmetric their contribution to the renormalization of ${\cal O}_{1,2}^{q}$ and
${\cal O}_{3,4}^{q}$ is proportional to $C_{1,2}^{q'q}+C_{1,2}^{qq'}$, which vanishes to great
accuracy in the present framework. One noticeable consequence is that the leading additive QCD
renormalization of the dipole operators $O_{3,4}^q$ comes at the NLO in the loop expansion (LO in
$\a_s$) from the ${\cal O}_{1,2}^{q'q}$ operators, in analogy to the $\Delta S=1$ case.

The QCD anomalous dimensions and mixings of the whole set of operators in \eq{eq:Ledm}, at the leading
order in the loop expansion, are found in Ref. \cite{Hisano:2012cc}.  A more recent NLO calculation
is presented in Ref. \cite{Brod:2018pli}. We report the relevant anomalous dimension matrix in
\app{app:anomalousdim} in our normalization. At the hadronic scale the Wilson coefficients
$C_1^{qq'}$ and $C_2^{qq'}$ turn out to be comparable, with a slight predominance of the
radiatively induced $C_1^{qq'}$.

\smallskip

The quark EDM from ${\cal O}_{3}^{q}$ is given in units of $e$ by
\beq
\label{qEDM}
d_q = -\frac{G_F}{\sqrt{2}}\frac{e_q\,m_q}{4\pi^2}\, C_{3}^{q}
\eeq
and analogously for the chromo-EDM
\beq
\label{qCEDM}
\tilde d_q = -\frac{G_F}{\sqrt{2}}\frac{m_q}{4\pi^2}\, C_{4}^{q}\, .
\eeq
The neutron EDM can be obtained from \eqs{qEDM}{qCEDM}, evaluated at the hadronic scale, via naive
dimensional analysis, chiral perturbation theory or QCD sum rules, the latter providing a more
systematic approach.  A recent reevaluation in this framework gives~\cite{Hisano:2012sc,Hisano:2012cc}~\footnote{
A chiral perturbation calculation of the chromoelectric dipoles gives coefficients larger by order one factors~\cite{Fuyuto:2012yf}, while a very recent lattice calculation gives~\cite{Yamanaka:2018uud} $d_n = 0.8d_d -0.2d_u$, still missing the chromoeletric dipoles. Since, as we shall see, the dipole contributions to the $n$EDM are largely subdominant, the present variance in the calculations is immaterial for our conclusions.
}
\beq
 d_n \simeq
 0.32d_d -0.08d_u +
e (0.12\tilde{d}_d-0.12\tilde{d}_u-0.006\tilde{d}_s) \,.
\label{QCDsumrules}
\eeq

\smallskip

In the presence of a PQ axion, the effective CP and chiral-symmetry breaking operators in
\eq{eq:Ledm} still induce a nonzero $\bar\theta$~\cite{Shifman:1979if,An:2009zh}, as explicitly derived in
\eq{PQvev} for ${\cal O}_1^{ud}$.  As a result, in the PQ case \eq{QCDsumrules} is modified to include the contribution
to the $n$EDM of the $\bar\theta$ induced by the chromoelectric dipoles~\cite{Pospelov:2000bw}
\beq d_n^{PQ} \simeq
 0.32d_d -0.08d_u +
e (0.25\tilde{d}_d+0.14\tilde{d}_u) \, . 
\label{QCDsumrulesPQ}
\eeq
The size of the $n$EDM induced by the $\bar\theta$ term is estimated by various methods to
be~\cite{Crewther:1979pi,Pich:1991fq,Pospelov:1999ha,Pospelov:1999mv,Shindler:2015aqa,Guo:2015tla,Chupp:2017rkp}
\beq
 d_n \simeq
-(\text{1--4})\times 10^{-16} \,\bar{\theta} 
\label{ThetaEDMn}
\eeq
in units of $e\cdot {\rm cm}$.
\smallskip

The contribution of the Weinberg three-gluon operator to the $n$EDM is subject to large hadronic
uncertainties, related to the method of evaluation.  By comparing different calculations one
finds~\cite{Demir:2002gg}
\beq
\label{GGG$n$EDM}
 d_n = -(\text{10--30}\ {\rm MeV}) \frac{G_F}{\sqrt{2}}\frac{e g_s}{8\pi^2}\, C_{5}(1\, {\rm GeV})\,.
\eeq
In the LR framework the dominant chirality flip of the dipole operators depends on the fermion
masses in the loop. Albeit chirally unsuppressed, the two-loop Weinberg operator ${\cal O}_5$ turns
out to give a subleading contribution to the $n$EDM.

\subsection{The long-distance contributions}

\noindent
The operator ${\cal O}_{1}^{qq'}$ mediates meson to vacuum transitions that when chirally rotated
away generate P and CP-violating interactions among mesons and baryons. These couplings induce potentially
large contributions to the $n$EDM via chiral loops~\cite{Khatsimovsky:1987fr,He:1992db}.  As shown in
Ref.~\cite{Haba:2018byj}, the pion VEV carries an enhancement factor $m_s/(m_u+m_d)$ with respect to
the other VEVs.  In the pion-baryon couplings it then dominates the chiral loop contributions
to the $n$EDM. As a matter of fact, by considering the $U(3)_L\times U(3)_R$ chiral Lagrangian
(Appendix~\ref{app:Lchiral}) with the inclusion of the axial anomaly term~\cite{Pich:1991fq} one
obtains
\beq
\vev{\pi^0} \simeq
\frac{G_F}{\sqrt{2}}({\cal C}_{1ud}-{\cal C}_{1du})\frac{4\, c_3}{B_0 F_\pi (m_d+m_u)}\,,
\label{pivevapp}
\eeq
with $\vev{\pi^0}\gg \vev{\eta_{0,8}}$ by a factor $m_s/(m_d-m_u)$. For the notation and estimate of
chiral couplings and low energy constants (LEC) see Appendix~\ref{app:Lchiral}.

Given the leading role of $\vev{\pi^0}$, the relevant CP violating baryon-meson couplings are (\eqs{nppi}{nsk})
\bea
\label{nppiapprox}
\bar{g}_{np\pi}\simeq&& \frac{2\sqrt{2} B_0}{F_\pi^2}(b_D+b_F) (m_d-m_u)\langle\pi^0\rangle \, ,
 \\[1ex]
\label{nskapprox}
\bar{g}_{n\Sigma^-K^+}=&& \frac{B_0}{F_\pi}(b_D-b_F)\left[
-2\sqrt{2}\ m_u\frac{\langle\pi^0\rangle}{F_\pi}\right.
 \\[0ex]
&&\left. -\frac{2\sqrt{2}}{\sqrt{3}}(m_u\!-\!2m_s)\frac{\langle\eta_8\rangle}{F_\pi}
-\frac{4}{\sqrt{3}}(m_u\!+\!m_s)\frac{\langle\eta_0\rangle}{F_0}\right] ,\nn
\eea
with $b_D+b_F\simeq -0.14$ and $b_D-b_F\simeq 0.28$ in units of $\rm GeV^{-1}$.  As we comment in
\app{app:Lchiral}, these results differ from those of Ref.~\cite{Haba:2018byj}.  In
particular, the coefficient of the pion VEV in \eq{nskapprox} is not enhanced by $m_s$. As a
consequence $\bar{g}_{n\Sigma^-K^+}$ is of size comparable to $\bar{g}_{np\pi}$ and all meson VEVs
are relevant.  In addition, when considering the PQ scenario, the $\bar\theta$ induced by the
${\cal O}_1^{ud}$ operator cancels exactly the $\bar{g}_{np\pi}$ coupling, leaving only
$\bar{g}_{n\Sigma^-K^+}$; the logarithmic enhancement in the pion mediated loop (visible below) is
therefore lost and the predicted $n$EDM is strongly suppressed.  On the contrary the
${\cal O}_1^{us}$ operator consistently induces a cancelation of $\bar{g}_{n\Sigma^-K^+}$ but a nonzero $g_{np\pi}$, 
so that, in spite of being doubly Cabibbo suppressed, the logarithmic enhancement of the pion
loop makes its contribution to the $n$EDM no longer negligeable, albeit still subdominant.  We
quantitatively discuss the effects of the pion coupling cancelation in the PQ setup in the following
section.

Up to unknown LECs (that are numerically subleading according to a naive dimensional estimate~\cite{Haba:2018byj}) the $n$EDM computed from baryon-meson chiral loops leads to~\cite{Ottnad:2009jw,Guo:2012vf}
\beq
d_n\simeq\frac{e}{8\pi^2F_\pi}\left.\frac{\bar{g}_{n\Sigma^-K^+}}{\sqrt{2}}(D-F)\left(\log\frac{m_K^2}{m_N^2}-\frac{\pi m_K}{2m_N}
\right)\right. ,
\label{dn_sigma}
\eeq
to be compared with the LO pion contribution~\cite{Seng:2014pba}
\beq
d_n\simeq -\frac{e}{8\pi^2F_\pi}\left.
\frac{\bar{g}_{np\pi}}{\sqrt{2}}(D+F)\left(\log\frac{m_\pi^2}{m_N^2}-\frac{\pi m_\pi}{2m_N}\right)\right. ,
\label{dn_pi}
\eeq
where at the leading order $D+F\equiv g_A\simeq 1.3$ and $D-F\simeq 0.3$. In
Ref.~\cite{Jenkins:1991es} large logarithmic corrections to the tree level result are computed
leading to $D+F\simeq 1$ and $D-F\simeq 0.2$. We include this spread within the hadronic uncertainty
in our numerical analysis. In \eqs{dn_sigma}{dn_pi} the extended-on-mass-shell prescription is
applied to ensure a correct power counting~\cite{Fuchs:2003qc}.

In spite of the large pion log in \eq{dn_pi}, \eq{dn_sigma} gives a non negligible contribution to the $n$EDM and we
will include it in our numerical analysis, together with the pion loop
contributions induced by $\bar g_{nn\pi}$~\cite{Seng:2014pba,Maiezza:2014ala,Cirigliano:2016yhc}.~\footnote{The direct short-distance contribution to the isovector CP odd pion nucleon coupling, which is part of the unknown loop counterterm, is estimated to be sizeable, albeit with a large uncertainty~\cite{Yamanaka:2017mef}. We assume no large cancellation occurs.}

\section{Numerical analysis}
\label{sec:numericaldiscussion}

\subsection{Preliminaries}

\noindent
According to \eq{hepsprime} and to the discussion of the LR hadronic matrix elements in
\sect{matrixelements} we obtain
\bea
h_{\eps'} &=& 0.92\!\times\!10^{6}\, |\zeta| \, \Big[ \sin \left(\a
   \!-\!\theta _u\!-\!\theta _d\right)+ \sin
   \left(\a\!-\!\theta_u\!-\!\theta_s\right) \Big] \nn\\
&&{} +\, 320\, |\zeta|\, \Big[ \sin \left(\a
   \!-\!\theta _c\!-\!\theta _d\right)+\sin \left(\a\!-\!\theta _c\!-\!\theta
   _s\right)  \Big]\nn \\
&&{}+6200\, r\,  \sin \left(\theta _d - \theta_s\right),
\label{eq:hep}
\eea
which is normalized to unity when matching the central experimental value in \eq{epspexp}.
The contributions that are proportional to the LR mixing $\zeta$ in \eq{mix} are due to
current-current (first line) and dipole operators (second line), while the term proportional to $r$
represents the RR current contribution. The relative magnitude of the three contributions is readily
estimated. From the direct search limit $M_{W_R}\gtrsim 3.7\,\TeV$ (see~\cite{Nemevsek:2018bbt}) one
obtains an upper bound on the mixing, $\zeta < 4\times 10^{-4}$. Thus, the first line can easily
overshoot by orders of magnitudes the experimental value. The contribution of the dipoles instead
amounts at most to $h_{\eps'} \sim 0.25$. As for the last line, the phase $\theta_d-\theta_s$ is
constrained by $\eps_K$, with different outcomes in the $\C$ or $\P$ cases. In the case of $\C$,
from \eq{eq:epsC} one finds $r\sin \left(\theta _d-\theta _s\right)<1.4\times 10^{-6}$, so that this
contribution falls short of $\sim 0.008$ and can be neglected. In the case of $\P$, from
\eq{eq:epsP} one finds
$r\sin\left(\theta _d-\theta _s\right)\lesssim 2.5\times 10^{-5}\,(7\,\TeV/M{_{W_R}})^{1.5}$, which
contributes to $h_{\eps'}$ at most as 0.15, where $M_{W_R}$ is constrained to be larger than
$7\,\TeV$ as we shall see in the following.

It is convenient to note that in the dominant contribution to $\eps'$, namely in
\beq
\label{eq:epspsimple}
h_{\eps'} \simeq 0.92\times 10^{6}  |\zeta|\Big[\!\sin \left(\a
   -\theta _u-\theta _d\right)+ \sin
   \left(\a -\theta _u-\theta
     _s\right)\! \Big],
\eeq
the constraints (\ref{eq:epsC})--(\ref{eq:epsP}) on $\theta _d-\theta _s$ enforced for low
LR scale imply that the result depends on a single combination of phases, e.g.\
$\a -\theta _u-\theta _d$.

\medskip

Turning to the neutron electric dipole moment, in analogy with $\eps'$ it is convenient to introduce the parameter
\begin{equation}
h_{d_n} =  \frac{|d_{n}^{LR}|}{d_{n}^{<}}\, ,
\end{equation}
where the LR contribution to the dipole moment is normalized to the present experimental bound,
$d_n^{<}=2.9\times 10^{-26}\,$e\,cm.

\begin{figure*}[t]
\centerline{\includegraphics[width=0.48\textwidth]{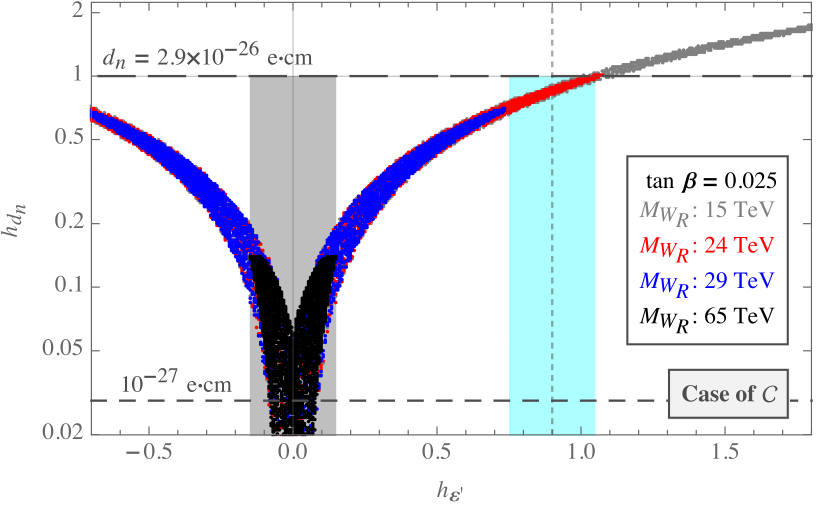}~\includegraphics[width=0.48\textwidth]{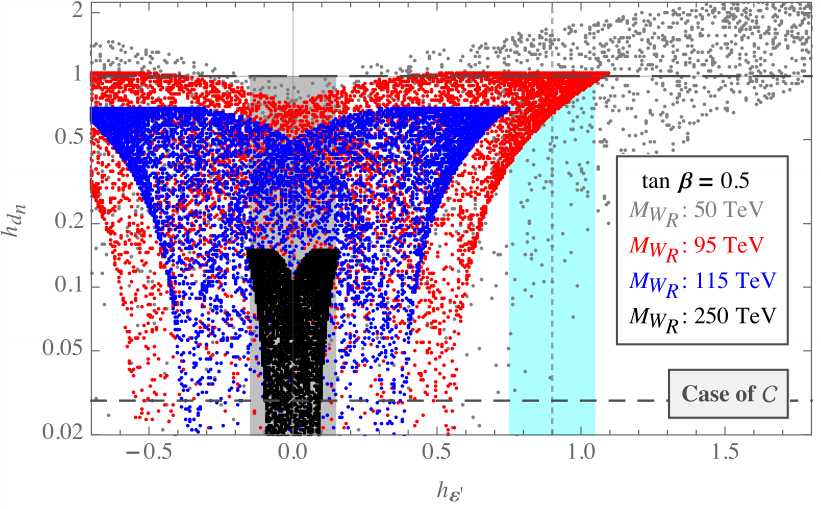}}
\caption{Case of $\C$: Contribution of the LRSM to $\eps'$ and $d_n$ for random phases, various choices of $M_{W_R}$ and two choices of $t_\b$: small $t_\b=m_b/m_t$ (left) or large $t_\b=0.5$ (right). Only those points that satisfy the $\eps$ constraint are shown. The left (gray) and right (cyan) vertical bands define the \SM and \NP scenarios respectively. The long-dashed line represent the current $d_n$ experimental bound and the short-dashed one a future reach of $d_n< 10^{-27} e\cdot{\rm cm}$.
\label{fig:scatter_C}}
\end{figure*}

From the results of the previous section and those in the appendices we finally obtain, for central
values of the hadronic parameters at the neutron scale,
\bea
h_{d_n}^{\text{noPQ}}&= 10^6  |\zeta|\,\Big|&
    {}+1.65 \sin \left(\a -\theta _u-\theta_d\right) \nn\\
  &&{}-0.007 \sin \left(\a -\theta _c-\theta_d\right)\nn\\
  &&{}+  0.00095 \sin \left(\a -\theta _t-\theta_b\right)\Big| \, ,
  \label{eq:dnC}\\
%
h_{d_n}^{\text{PQ}}&= 10^6  |\zeta|\,\Big|&
     {}+0.21 \sin \left(\a -\theta _u-\theta_d\right) \nn\\
  &&{}-0.010 \sin \left(\a -\theta _c-\theta_d\right)\nn\\
  &&{}+  0.00095 \sin \left(\a -\theta _t-\theta_b\right)\Big| \, ,
   \label{eq:dnPQ}
\eea
where the first line includes the contribution of the LR current-current operators via chiral loops,
generated by the $\bar{g}_{np\pi}$, $\bar{g}_{n\Sigma^-K^+}$ and $\bar g_{nn\pi}$ couplings,
and in the PQ case it includes the induced shift on $\bar\theta$ (see \eq{PQvev}). As already mentioned in the previous section,
the ${\cal O}_1^{ud}$ induced $\bar\theta$ cancels exactly the $\bar{g}_{np\pi}$ coupling, leaving only $\bar{g}_{n\Sigma^-K^+}$ to contribute in chiral loops to the $n$EDM (see the discussion in \app{app:Lchiral}). The logarithmic enhancement
of the pion mediated loop is therefore lost with a consequent suppression of the predicted $n$EDM.

The much smaller second and third lines derive from the dipole and the Weinberg operators
respectively (including the renormalization mixing). The results in the noPQ case are in fairly good
agreement with those reported in Ref.~\cite{Haba:2018byj}.

In the discussion that follows we will consider these outcomes as benchmark values. In order to
consider the uncertainties discussed in the previous sections, we allow a range of 50\%--200\% for
$h_{\eps'}$ due mostly to the relevant LR matrix elements, and a 30\% uncertainty on $h_{d_n}$
related to the long-distance parameters $D$ and $F$ in \eqs{dn_sigma}{dn_pi}.

\medskip

It appears immediately that the combinations of phases in the leading terms of \eq{eq:hep} and
\eqs{eq:dnC}{eq:dnPQ} can lead to correlations that open the possibility to test the LR setup.  This
is especially clear for low scale $M_{W_R}< 30\,$\TeV\ because, in view of the $\eps$ constraints in
\eqs{eq:epsC}{eq:epsP}, $\theta_d$ and $\theta_s$ are strongly related.

In the following, we explore this correlation by considering the two phenomenological scenarios for
$\eps'/\eps$: namely, whether the SM prediction falls short of the experimental value with the
missing contribution being provided by low scale LRSM, or whether the SM prediction saturates the
observable and thus a lower bound on the $W_R$ mass follows.

\medskip

It is then crucial to consider the difference between the $\P$ and $\C$ choice of the discrete LR
symmetry. The important feature is that the phases $\theta_i$ are free for $\C$, while for $\P$ they
are predicted as a function of $s_\a t_{2\b} $. Therefore, in the case of $\C$ one can always
suppress the effects CP violation by setting the phases to zero, and thus no lower bound can be
placed on the $W_R$ scale. On the contrary, requiring a sizeable contribution to $\eps'$ bounds the
size LR scale from above. For the case of $\P$ the phases can be calculated analytically in a power
series of $s_\a t_{2\b}$, as demonstrated in~\cite{Senjanovic:2016vxw}. For our purposes it is
enough to consider the leading order expressions, which we recalculate for generic signs $s_i$ (see
Appendix~\ref{app:phases} for the detailed expressions).

\medskip

In summary, for the sake of clarity, we shall discuss our results with reference to two alternative
scenarios in which i) the SM saturates the experimental value of $\eps'/\eps$ and thus the LRSM
contribution has to be bounded from above, or ii) the SM contribution falls within the present
experimental error and the LRSM contribution provides about the whole amount. We name these limiting
scenarios \SM and \NP respectively. In either case we require the contribution to $d_n$ to be less
than the present experimental bound (the effect of future experimental improvements are also shown).
Thus, we set

\medskip

\qquad\quad \SM : \quad\  $h_{d_n}<1.$ \  and \ $| h_{\eps'}| <0.15$\,,

\smallskip

\qquad\quad  \NP : \quad\  $h_{d_n}<1.$ \  and \ $h_{\eps'}=0.9\pm0.15$\,.

\medskip

\noindent
The  uncertainty of 0.15 is related to the present experimental error on $\eps'/\eps$.

A discussion of the correlated predictions for $h_{\eps}$, $h_{\eps'}$ and $h_{d_n}$ will be given
next in the cases of $\C$ and $\P$, distinguishing between the \SM and \NP scenarios.

\subsection{Results}

\noindent
Our results are summarized in Figures 1 to 5. 
We analyze separately the $\C$ and $\P$ cases. The scattered plots are obtained for benchmark values of the hadronic parameters.

\begin{figure}
  \vspace*{-1ex}
  \centerline{\includegraphics[width=1\columnwidth]{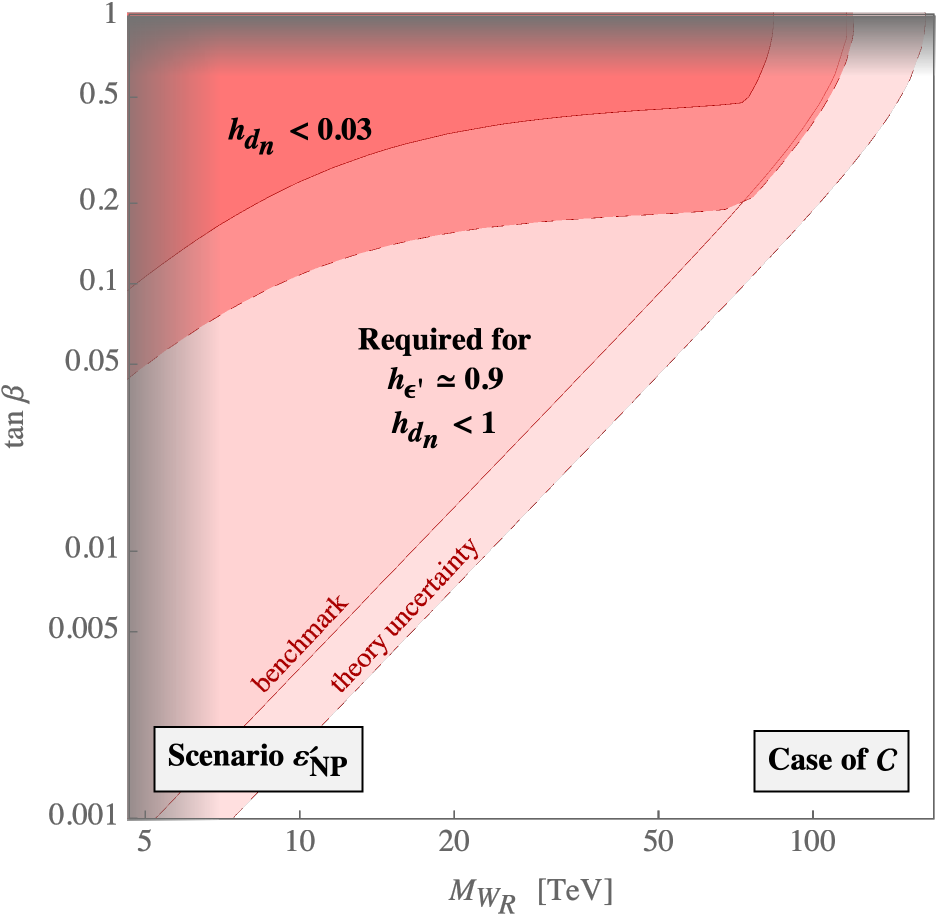}}
  \vspace*{-2.3ex}
  \caption{Case of $\C$ in \NP scenario: Allowed region in the $M_{W_R}$--\ $t_\b$ plane for
  $0.9\ \eps'_{\text exp}$ to arise from new physics. The $n$EDM prediction is taken below 1 (light shading) or 0.03 (dark shading) of the present experimental
  bound. Solid lines correspond to the chosen benchmark values of the hadronic parameters, while the
  dashed contours include the theoretical uncertainty. In the progressively shaded band at the top,
  $t_\b\gtrsim 0.6$, quark Yukawa couplings become non-perturbative~\cite{Maiezza:2010ic}, while at the left, for
  $M_{W_R}< 6\,\TeV$, the scalar sector becomes non-perturbative~\cite{Maiezza:2016bzp,Maiezza:2016ybz,Chauhan:2018uuy}.\label{fig:masterC}}
\vspace*{-2.5ex}
\end{figure}

\medskip

{\bf Case of $\C$.} Because the phases are free parameters, the LRSM contribution to both $\eps'$
and $d_n$ can be made vanishing by appropriate tuning.  Correspondingly, in \fig{fig:scatter_C} the
dots populate the gray band around zero no matter how low $M_{W_R}$ is.  As a result, no bound on $M_{W_R}$ can
be placed in the \SM scenario.

In the \NP scenario instead, because one requires a sizeable contribution of the LRSM to $\eps'$, an
upper bound on $M_{W_R}$ appears. Its size clearly depends on $t_\b$. For instance, for large
$t_\b\simeq 0.5$, near its perturbativity limit (right frame in \fig{fig:scatter_C}), one sees that
$W_R$ must be lighter than $115\,\TeV$, while for $t_\b=0.025$, $M_{W_R} < 30\,\TeV$ is required
(left frame).

One also notices that the correlation between $\eps'$ and $d_n$ is sharper (thinner band) for $W_R$
lighter than $\sim 30$ TeV and small $t_\b$ (left frame). In this regime due to the $\eps_K$
constraint $\t_s-\t_d$ has to be small (modulo $\pi$); the dominant first lines in
Eqs.~(\ref{eq:dnC}) and (\ref{eq:hep}) depend on the same phase combination and are thus confined in
a tiny strip.  This correlation is progressively absent in the large $t_\b$ regime (right frame)
because there $W_R$ is much heavier and the $\eps$ constraint becomes less effective.

In any case, since the free phases as well as $\a$ are at present not directly tested by other
observations, it is convenient to marginalize over them and show the resulting upper bound on
$M_{W_R}$ correlated with $t_\b$. This is depicted in \fig{fig:masterC}, where the upper bound on
$M_{W_R}$ in the \NP scenario is seen to range from less than 10\,\TeV\ (for $t_\b\sim 10^{-3}$) up
to more than 100\,\TeV\ (for large $t_\b$).

In \fig{fig:masterC} we also show the effect of tightening the constraint on $d_n$ to $<0.03$, in
view of the future experiments. One can conclude that when this bound will be reached, the LRSM
contribution to $\eps'$ can take place only for somewhat large $t_\b\gtrsim 0.1$.

We depict with a dashed contour the bound obtained after including the theoretical uncertainties dominated by
those on the matrix elements.  A numerical difference can be appreciated, but the picture patterns
remain.

\medskip

\begin{figure}
  \centerline{\includegraphics[width=0.98\columnwidth]{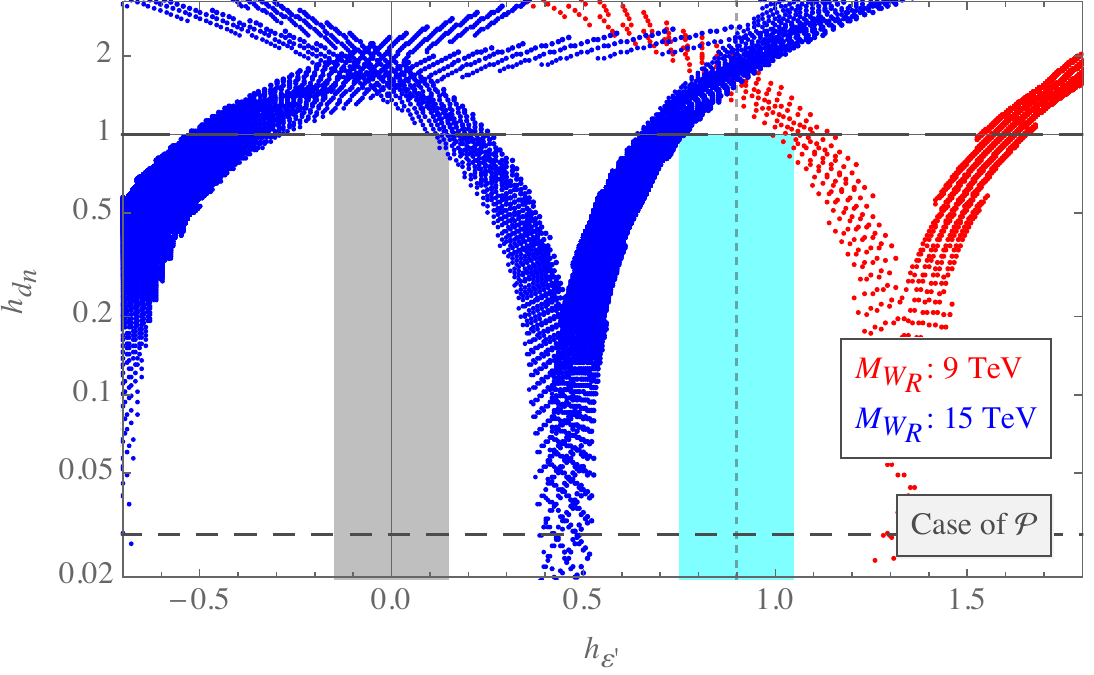}}
  \vspace*{-2.3ex}
  \caption{Case of $\P$: Contribution of the LRSM to $\eps'$ and $d_n$ for various choices of $M_{W_R}$ by scanning on phases and $t_\b$, while satisfying the $\eps$ constraint.  The left (gray) and right (cyan) vertical bands define the \SM and \NP scenarios respectively. The long-dashed line denotes the current experimental upper bound, while the short-dashed one a future reach of $d_n< 10^{-27} e\cdot{\rm cm}$.
    \label{fig:scatter_P}}
  \vspace*{-2.5ex}
\end{figure}

{\bf Case of $\P$.} In this case all phases are predicted in terms of $\a$ and $t_\b$, and although
the 32 different combinations of signs $s_{u,c,t,d,s}$ (see \eq{phasesigns}) give rise to different
numerical predictions, the resulting picture illustrated in \fig{fig:scatter_P} shows well defined
and narrow bands.  One can observe that the new physics contribution
to $\eps'$ shows a different pattern with respect to the previous case.

First, for low scale $W_R$ it is not possible to make $\eps'$ vanishing by a convenient choice of
phases, and thus a lower bound on $M_{W_R}$ emerges in the \SM scenario. As anticipated above, the
reason is the role played by the $\eps_K$ constraint~\eq{eq:epsP}: for low scale $W_R$ one must have
quite a large $\t_s-\t_d\sim\pm 0.16$~\cite{Bertolini:2014sua} and so the combinations in
\eq{eq:hep} can never vanish in correspondence of vanishing $h_{\eps'}$, $h_{d_n}$.  Thus, by
lowering $M_{W_R}$ the predicted values of $h_{\eps'}$ shift to larger values, as shown in
\fig{fig:scatter_P}.

From the plot it can be argued that the \SM scenario requires $M_{W_R}\gtrsim 15\,\TeV$.  The
detailed analysis gives information on the correlations between the phases and scales involved, as
we report in appendix~\ref{app:Pplots}, that shows the complex interplay of the 32 sign combinations
and the tight correlation among $\alpha$, $t_\b$ and $M_{W_R}$.

\begin{figure}[t]
  \centerline{\includegraphics[width=1\columnwidth]{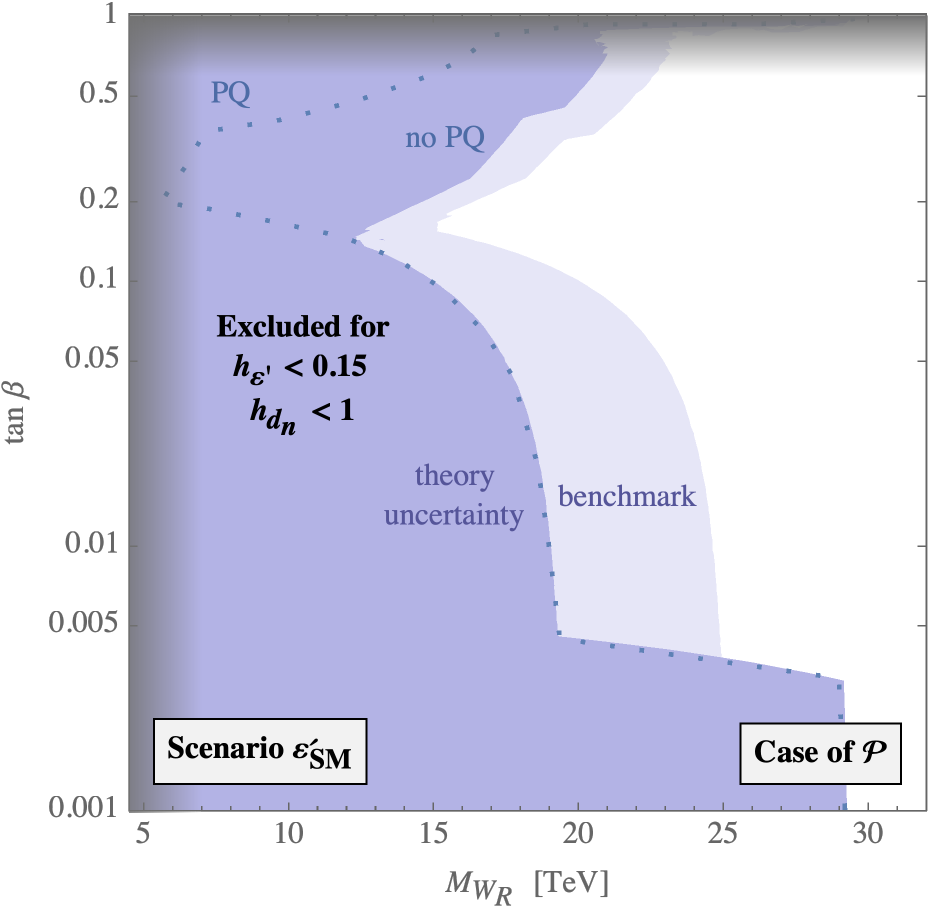}}
    \vspace*{-1.ex}
    \caption{Case of $\P$: The shaded regions in the $M_{W_R}$--$t_\b$ plane are excluded in order to have
      at most 15\% new physics contribution to $\eps'/\eps$ and $d_n$
      below the present experimental bound. The lighter region assumes benchmark 
      hadronic parameters, while the darker one includes the theory uncertainties,
      as discussed in the text. The dotted line marks the PQ case.
      \label{fig:masterPbound}}
    \vspace*{-.5ex}
\end{figure}

With the numerical study at hand, by marginalizing on $\a$ we depict in~\fig{fig:masterPbound} the
lower bound on $M_{W_R}$ as a function of $t_\b$,.  We find that the lowest allowed scale
$M_{W_R}\gtrsim 15\,\TeV$ is achieved for $t_\b\sim 0.15$.
The darker area shows the impact of including the theory uncertainty dominated by the hadronic
matrix elements (100\% for $\eps'$ and 30\% for $n$EDM).  This relaxes the lower bound to
a least possible value of $M_{W_R}\gtrsim 13\,\TeV$, which is achieved for $t_\b\sim 0.1$.
In the PQ scenario (dotted line) the lower bound is relaxed to about $6\,\TeV$ for $t_\b\sim 0.2$.

\smallskip

\begin{figure}[t]
  \centerline{\includegraphics[width=1\columnwidth]{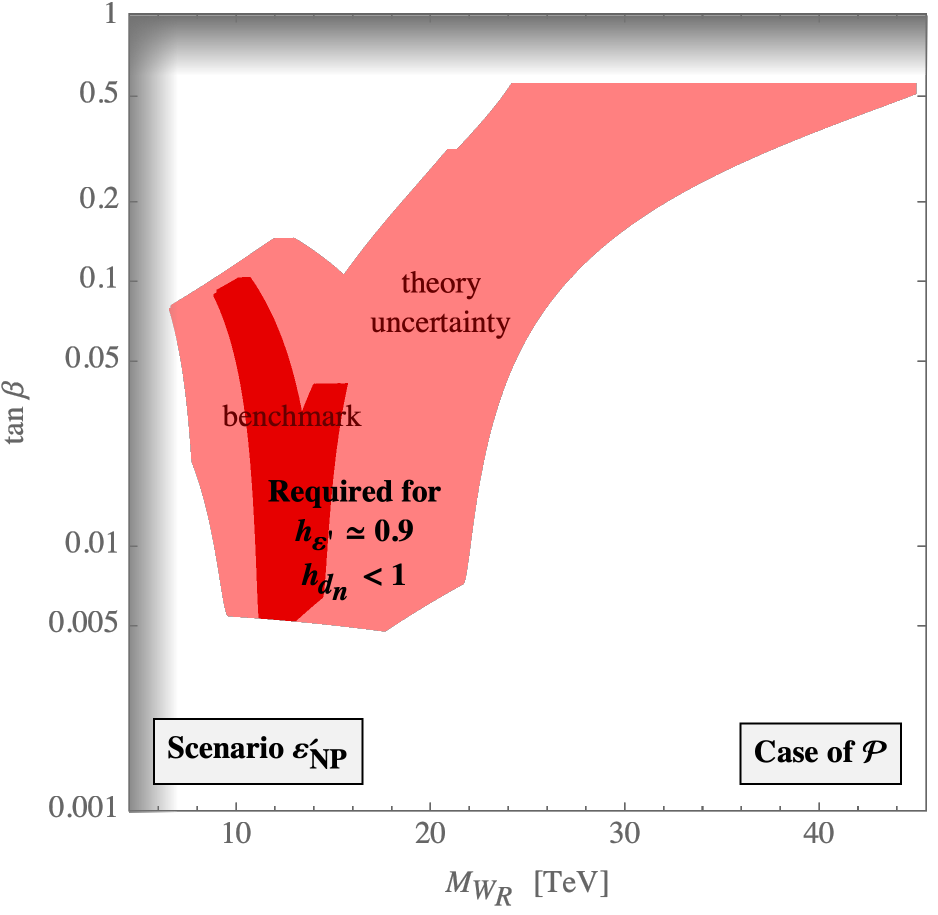}}
  \vspace*{-1.ex}
  \caption{Case of $\P$: The shaded regions in the $M_{W_R}$ -- $t_\b$ plane are allowed in order for new
    physics to provide $0.9\ \eps'_{\text exp}$, while keeping the contribution to $d_n$ below the present
    experimental bound. The smaller region assumes benchmark values for the hadronic parameters, while the
    larger one includes the theory uncertainties, as discussed in the text.  
    \label{fig:masterPnp}}
  \vspace*{-.5ex}
\end{figure}

In the \NP case the situation is even more structured and interesting. The constraint $h_{d_n}<1.$
can be satisfied by choosing the vacuum phase $\a$ appropriately, if possible, thus providing a
prediction of $h_{\eps'}$.  As expected, the request $h_{\eps'}\simeq 0.9\pm 0.15$ sharply
constrains the range of $M_{W_R}$. For instance in \fig{fig:scatter_P}, for benchmark values of the
parameters, a preferred range of $M_{W_R}=9$--15$\,\TeV$ emerges. This is better seen in
\fig{fig:masterPnp} where by marginalizing again on $\a$ we depict the allowed region in the plane
$M_{W_R}$--$t_\b$ for the \NP scenario in the case of $\P$ with benchmark hadronic parameters
(darker shaded area).  The region spans the perturbative interval $t_\b\simeq 0.005$--0.1 and
$M_{W_R}$ below 15\,\TeV. When we include the hadronic uncertainties (lighter shaded area) we find
that the allowed region relaxes substantially in the $M_{W_R}$ direction, which spans the
$7$--$45\,\TeV$ interval.

\pagebreak[3]

In summary, in the case of $\P$, our numerical analysis shows that with conservative theoretical
uncertainties the \SM scenario places a lower bound $M_{W_R}\gtrsim 13\,\TeV$, while the \NP
scenario requires $M_{W_R}=7$--$45\,\TeV$ and $t_\b=0.005$--0.5. Both these bounds will get
considerably tighter with the expected reduction of the theoretical errors on $\eps'$, coming from a
precise lattice determination of the relevant hadronic matrix elements.

\vspace*{-1ex}

\section{Conclusions}
\label{conclusions}

\noindent
In this work we have revisited the CP-violating observables $\eps$, $\eps'$ and $d_n$ within the
LRSM and derived updated limits on the LR scale $M_{W_R}$ and model parameters. In particular, our
analysis aimed to fully investigate the correlations among the observables for the different setups
of the model parameters, arising by a given choice of discrete LR symmetry.

The issue is relevant also in the light of the recently reopened debate on the magnitude of the
theoretical SM prediction for $\eps'$ following the recent lattice results, suggesting that the SM
may fall short of reproducing the experimental value of $\eps'/\eps$.  While waiting for a fully
consistent picture of the kaon hadronic decays from lattice, we considered here two limiting
scenarios named as \SM and \NP respectively, according to whether the SM or the LRSM saturate the
experimental value.

In the LRSM each choice of $\P$ or $\C$ discrete symmetry implies crucially different predictions
and constraints for the RH CKM phases with a strong impact on the new physics scale.

\smallskip

Previous detailed studies on $\eps,\eps'$ in the LRSM can be found
in~\cite{Bertolini:2012pu,Bertolini:2013noa,Bertolini:2014sua}, and of $\eps$, $\eps'$, $d_n$
in~\cite{Maiezza:2014ala,Haba:2018byj}. In particular, the LRSM prediction for $\eps$ was thoroughly
analyzed in~\cite{Bertolini:2014sua} and gives a tight constraint on the RH CKM phases versus the LR
scale.  On the other hand the relation with $d_n$ was studied recently in \cite{Maiezza:2014ala}
mainly within the assumption of exact $\P$ parity and thus with vanishing QCD theta term (although
the possibility of a different UV completion was foreseen).

\smallskip

In order to investigate an interesting and predictive scenario for the neutron dipole moment, we
considered explicitly the PQ setup as a dynamical solution of the theta QCD problem and investigated
in detail the impact of the explicit breaking of the PQ symmetry by the effective operators in the
minimal LRSM.  Our analysis shows that the new contribution to $d_n$ coming from the shift of the
axion potential, has a dramatic impact on the outcome, suppressing substantially the predicted $n$EDM.
This result is at variance with the conclusions of Ref.~\cite{Haba:2018byj}, and we identify the
issue in the proper extraction of the relevant meson-baryon couplings from the $U(3)$ chiral
lagrangian.  The inclusion of the strange quark in the analysis does not provide the large
enhancement of the chiral loops claimed in~\cite{Haba:2018byj} but still it is numerically relevant
and we evaluate the chiral loop contributions to $d_n$ within the $U(3)$ chiral framework (not
considered in~\cite{Maiezza:2014ala}).

Other studies on this topics have recently appeared in the literature. In particular i)
in~\cite{Dekens:2017hyc} the authors address the RH interactions in an effective parametrization,
ii) in~\cite{Haba:2018byj} the authors address the LRSM model with $\C$ symmetry, seemingly ignoring
the phase correlations deriving from the $\eps$ constraint, and focusing on small $t_\b$ only; iii)
in~\cite{Haba:2017jgf} the case of $\P$ symmetry is analyzed by decoupling the scale of LR gauge
interactions, $M_{W_R}$, and effectively studying a two Higgs doublet model.  The study presented
here aims to provide a comparative picture of the implications of $\eps$, $\eps'$ and $d_n$ on the
minimal LRSM setup, exhibiting the main patterns of the model
parameters and scales due to the correlations among the observables.

\smallskip
Given the suppression of the $n$EDM in the effective PQ scenario, which does not lead to significant bounds on the LR scale, we analyzed in detail, for both $\C$ and $\P$, the case
where the ultraviolet completion of the low energy minimal LR model sets $\bar\theta =0$ (see for instance~\cite{Barr:1991qx}). 
This leads to interesting and predictive scenarios for the $n$EDM. 

Our conclusions in the \SM case can be summarized as follows: for the choice of $\C$ symmetry,
which allows for several free phases, one finds constraints on them but no lower bound on the RH
scale is present. For the choice of $\P$ instead, where all the CKM phases are predicted in terms of one
vacuum phase, we find a lower bound $M_{W_R}>13\,\TeV$, which includes conservative theoretical uncertainties
on the hadronic parameters. Future improvement on these uncertainties will push this bound slightly
higher, up to $15\,\TeV$. This is in any case smaller than the tight lower bound of $\sim 30\,\TeV$
which stems from exact parity with spontaneously induced $\bar\theta\neq 0$ ~\cite{Maiezza:2014ala}.

In the \NP scenario, the LRSM contributions may saturate $\eps'$ and still hold the $d_n$ below the
experimental bound or at the reach of future probes. An upper limit on $M_{W_R}$ is naturally
demanded for this to happen.  The case of $\P$, being more constrained, requires
$M_{W_R}=7$--$45\,\TeV$ (or $M_{W_R}=9$--$15\,\TeV$ with benchmark hadronic parameters) and
$t_\b>0.005$. On the other hand for the $\C$ case, the LR contributions can saturate
$\eps'$ for $M_{W_R}$ ranging from about 8 to 115\,\TeV\ with increasing $t_\b$, from $0.001$ to its
perturbativity bound $\simeq 0.5$.


The scale of $M_{W_R}\sim 7\,\TeV$ lies just at the limiting reach of LHC in the golden KS
channel~\cite{Nemevsek:2018bbt}. On the other hand, it may show up as an effective interaction
either in the dilepton channel or, for a particular range of neutrino masses, in displaced decays of
the Higgs to two RH neutrinos~\cite{Maiezza:2015lza}.

A future hadronic collider at $30\,\TeV$ center-of-mass energy would probe easily the mass scale up
to $\sim$15\,TeV in the KS process~\cite{Ruiz:2017nip} (for TeV scale RH neutrinos) or in the lepton
plus missing energy channel~\cite{Nemevsek:2018bbt} (for RH neutrinos below 100\,\GeV), the two
having comparable reach. The prominent $W_R\to$ dijet channel would give an earlier signature,
independently from the RH neutrino mass.

Finally, a FCC-hh collider with 100\,\TeV\ center-of-mass energy would probe $M_{W_R}$ up
to $40\,\TeV$~\cite{Ruiz:2017nip,Chauhan:2018uuy} and thus test thoroughly the scenarios analyzed
here.


\section*{Acknowledgments}

\noindent
We thank Goran Senjanovi\'c for interesting discussions. AM is partially supported by the Croatian
Science Foundation project number 4418.

\appendix

\section{The external phases of $V_R$}
\label{app:phases}

\noindent
In the case of $\P$ the external phases can be worked out from the perturbatively computed
$V_R$~\cite{Senjanovic:2014pva} in~\eq{VRP}. For the analysis on $d_n$ and $\eps'$ performed in this
article, the relevant phase combinations are $\theta_u+\theta_d$ and $\theta_u+\theta_s$, that can
be expressed in terms of the quark masses, the CKM angles, the expansion parameter $s_\a t_{2\b}$
and the arbitrary signs $s_i$:

\begin{widetext}
  {\small
    \begin{align}\label{ud_us}
 & \theta_u+\theta_d \simeq
\frac{s_\a t_{2\b}}{2} \bigg[\sin ^2\!\theta _{12} \left(
   \frac{2s_s}{m_s}-\frac{s_d}{m_d}\right)\!\left(m_c s_c \cos ^2\!\theta _{23}+ m_t s_t\sin ^2\!\theta _{23}
   \right) - \frac{s_u}{m_u}\! \left(m_d s_d \cos ^2\!\theta _{12}+s_s  m_s \sin
   ^2\! \theta _{12}\right)\bigg]+\frac{s_u - s_d}2\pi\, ,  \\
  & \theta_u+\theta_s \simeq
 \frac{s_\a t_{2\b}}{2} \bigg[\cos ^2\!\theta _{12} \frac{s_s}{m_s}\! \left(m_c s_c
   \cos ^2\!\theta _{23}+ m_t s_t\sin ^2\!\theta _{23}\right)-\frac{s_u}{m_u}\!
   \left(m_d s_d \cos ^2\!\theta _{12}+m_ss_s\sin ^2\!\theta _{12}
   \right)\bigg]+\frac{s_u - s_s}2\pi\,.
    \end{align}
    }%
\end{widetext}
%

\section{Loop functions}
\label{app:loop}

\noindent
The loop functions relevant for the SM and the LRSM short-distance coefficients in
\eqs{eq:C2AB}{eq:Cg12} are given by~\cite{Ecker:1985vv,Inami:1980fz,Cho:1993zb,Buras:1998raa}
{\small
  \bea
&&F_1^{LL}=\frac{x(-18+11 x+x^2)}{12 (x-1)^3}-\frac{(4-16 x+9 x^2) \ln x}{6 (x-1)^4} \qquad 
\eea
}{\small
  \bea
&&E_1^{LL}=-\frac{x^2(5x^2-2x-6)}{18(x-1)^4}\ln x+\frac{19x^3-25x^2}{36(x-1)^3}+\frac{4}{9}\ln x\ \ \ \ \ \ \ \ \\  
&&F_2^{LL}=\frac{x(2+3x-6x^2+x^3+6x \ln x)}{4 (x-1)^4} \\[0ex]
&&E_2^{LL}=\frac{x (8x^2+5x-7)}{12(x-1)^3}+\frac{x^2(2-3 x)}{2(x-1)^4}\ln x  
\\ 
&&F_2^{LR}=\frac{-4+3x+x^3-6x \ln x}{2(x-1)^3} \\[0ex]
&&E_2^{LR}=\frac{5x^2-31x+20}{6(x-1)^2}-\frac{x(2-3 x)}{(x-1)^3}\ln x\,.
\eea
}%
One also has $F^{RR}_{1,2}=F^{LL}_{1,2}(r x_i)$ and similarly for $E_{1,2}^{RR}$.

     \section{Anomalous dimensions}
     \label{app:anomalousdim}

\noindent
As explained in Sect.~\ref{sec:neutronedm} the pattern of model values of the weak scale Wilson
coefficients in \eq{eq:Ledm} allow us to reduce the basis of the effective operators to
\bea
&&{\cal O}_{qq'} = \{{\cal O}_1^{qq'},\, {\cal O}_2^{qq'},\, {\cal O}_1^{q'q},\, {\cal O}_2^{q'q}\}\,, \\[1ex]
&&{\cal O}_q \ \  = \{{\cal O}_3^{q},\ {\cal O}_4^{q}\}\,.
\eea

The LO mixing matrix of the ${\cal O}_{qq'}$ operators is given by~\cite{Hisano:2012cc}
\begin{align}
\label{eq:qqptoqqp}
\gamma^{(1)}_{qq'\to qq'} &=
\begin{pmatrix}
-16        & 0                   & 0             & 0                 \\[0.3ex]
-6            & 2          & 0             & 0               \\[0.3ex]
0             & 0                   & -16        & 0              \\[0.3ex]
0             & 0                   & -6            & 2
\end{pmatrix},
\end{align}
while the dipole anomalous-dimension matrix (we neglect the mixing of ${\cal O}_{5}$ into ${\cal O}_4^q$)
reads~\cite{Misiak:1994zw,Degrassi:2005zd}
\begin{align}
\label{eq:dipolemix}
\gamma^{(1)}_{q\to q} &=
\begin{pmatrix}
\frac{32}{3} & 0         \\[0.3em]
\frac{32}{3} & \frac{28}{3}
\end{pmatrix}.
\end{align}
The subscripts $qq'$ and $q$ in the $\gamma$'s indicate the non-vanishing subblocks of the
anomalous-dimension matrix.

The mixing of ${\cal O}_{qq'}$ into the dipoles is readily obtained from Ref.~\cite{Brod:2018pli},
taking into account the different operator basis and the covariant derivative conventions
\begin{align}
\label{eq:qqptoq}
\gamma^{(1)}_{qq'\to q} &=
\begin{pmatrix}
   \frac{8}{3}\frac{e_{q'}}{e_q}\frac{m_{q'}}{m_q} & -\frac{67}{6}\frac{m_{q'}}{m_q} \\[0.5em]
  -8\frac{e_{q'}}{e_q}\frac{m_{q'}}{m_q} & -\frac{5}{2}\frac{m_{q'}}{m_q} \\[0.5em]
  \frac{8}{9}\frac{e_{q'}}{e_q}\frac{m_{q'}}{m_q} & -\frac{67}{18}\frac{m_{q'}}{m_q} \\[0.5em]
  -\frac{8}{3}\frac{e_{q'}}{e_q}\frac{m_{q'}}{m_q} & -\frac{5}{6}\frac{m_{q'}}{m_q}
\end{pmatrix}.
\end{align}

The Wilson coefficients evolve according to
\beq
\label{RGE}
\frac{{\rm d} C}{{\rm d} \log\mu} =  C \gamma^{(1)} \frac{\a_s}{4\pi}\,,
\eeq
where $\gamma^{(1)}$ is the $6\times 6$ anomalous dimension matrix. The $\gamma$'s upperscript
indicates the $\a_s/4\pi$ order of the mixing.

The short-distance running of the LR effective operators for $\Delta S=1$ and $\Delta S=2$
transitions is discussed in Refs.~\cite{Bertolini:2012pu,Bertolini:2014sua}, respectively.

\section{{The meson and baryon chiral Lagrangians}}
\label{app:Lchiral}

\noindent
The LO chiral Lagrangian for the octet of Nambu-Goldstone bosons and the $\eta_0$ singlet, including
the bosonic representation of the ${\cal O}_{1q'q}$ operator is given by
\bea
{\cal L}_{\rm }={}&&\frac{F_\pi^2}{4}{\rm tr}\left[(D_\mu U)^\dagger D^\mu U+\chi(U+U^\dagger)\right]
\nonumber\\[0ex]
&&+ a_0 \, {\rm tr}\left[\log U-\log U^\dagger\right]^2
\nonumber \\[1ex]
&&+ \frac{G_F}{\sqrt{2}}\sum_{u,d,s}\Big\{
i{\cal C}_{ijkl}^{LRLR}    \left(c_1 [U]_{ji}[U]_{lk}-c_1 [U^\dagger]_{ji}[U^\dagger]_{lk}   \right.
\nonumber \\[0ex]
&&\quad+ \left.c_2 [U]_{li}[U]_{jk}-c_2 [U^\dagger]_{li}[U^\dagger]_{jk}\right)
\nonumber \\[1.5ex]
&&\quad+i{\cal C}_{ijkl}^{RLLR} \left(c_3 [U^\dagger]_{ji}[U]_{lk}-c_3 [U]_{ji}[U^\dagger]_{lk}\right)\Big\}\,  ,\ \ 
\label{Lmeson}
\eea
where we follow the notation of Ref.~\cite{Haba:2018byj}.  The $3\times 3$ matrix $U$
represents nonlinearly the nine Goldstone states. Under $U(3)_L\times U(3)_R$ rotations $L\times R$
it transforms as $U\to RUL^\dagger$, while $\chi$ includes explicitly the quark masses, namely
\bea
&&U=\exp\left[\frac{2i}{\sqrt{6}F_0}\eta_0\,I+\frac{2i}{F_\pi}\Pi\right] ,
\\[4ex]
&&\Pi\equiv\begin{pmatrix} 
  \frac{1}{2}\pi^0+\frac{1}{2\sqrt{3}}\eta_8 & \frac{1}{\sqrt{2}}\pi^+ & \frac{1}{\sqrt{2}}K^+ \\
  \frac{1}{\sqrt{2}}\pi^-  &  -\frac{1}{2}\pi^0+\frac{1}{2\sqrt{3}}\eta_8 & \frac{1}{\sqrt{2}}K^0 \\
  \frac{1}{\sqrt{2}}K^- & \frac{1}{\sqrt{2}}\bar{K}^0 & -\frac{1}{\sqrt{3}}\eta_8 \\
   \end{pmatrix}, \ \ \ 
\\[4ex]
&&\chi=2 B_0 \ {\rm diag}\{m_u,m_d,m_s\} 
\eea
and $I$ is the identity matrix.  $F_\pi$ is the pion decay constant in the chiral limit, while $F_0$
denotes the $\eta_0$ decay constant, which we approximate to be equal. The quark mass term is
written in terms of the condensate $B_0\simeq m_\pi^2/(m_u+m_d)$.

The second term in \eq{Lmeson} represents the anomaly induced by the QCD instantons in the large $N$
limit~\cite{Pich:1991fq}.  The coupling $a_0$ satisfies
$48a_0/F_0^2\simeq m_\eta^2+m_{\eta'}^2-2m_K^2$.

The third term represents the bosonization of ${\cal C}_{1q'q}{\cal O}_{1q'q}$ where the sum over
${q\neq q'=u,d,s}$ is understood. The coefficients that encode the short distance physics are given
by
${\cal C}_{ijkl}^{LRLR}={\cal C}_{ijkl}^{RLLR}\equiv\sum_{q\neq q'}{\cal C}_{1q'q} \
\delta_{iq'}\delta_{jq'}\delta_{kq}\delta_{lq}$.  The unknown low energy constants (LEC)
$c_{1,2,3}$, are estimated in the large N limit as
\beq
c_1\sim c_2\sim c_3\sim \frac{F_\pi^4 B_0^2}{4}\,.
\label{c1c2c3}
\eeq

The terms proportional to $c_1$ and $c_3$ induce VEVs of the Goldstone nonet. However, the $c _1$
terms are proportional to $({\cal C}_{1ud}+{\cal C}_{1du})$, which vanishes due to \eq{C12ud-LR}.
Thus, only the $c_3$ contributions, proportional to $({\cal C}_{1ud}-{\cal C}_{1du})$ are non
vanishing. By neglecting $|{\cal C}_{1us}|$, doubly Cabibbo suppressed  with respect to $|{\cal C}_{1ud}|$, we
confirm the results in~\cite{Haba:2018byj}
\bea
\label{pivev}
\frac{\langle\pi^0\rangle}{F_\pi}\simeq {} &&
\frac{G_F}{\sqrt{2}}({\cal C}_{1ud}-{\cal C}_{1du})\frac{c_3}{B_0F_\pi^2}  \\
&&\!\!\!\!\times \frac{B_0F_\pi^2(m_u+m_d)m_s+8a_0(m_u+m_d+4m_s)}{B_0F_\pi^2m_um_dm_s+8a_0(m_um_d+m_dm_s+m_sm_u)},
 \nn \\[1.5ex]
\label{eta8vev}
\frac{\langle\eta_8\rangle}{F_\pi}\simeq {} &&
\frac{G_F}{\sqrt{2}}({\cal C}_{1ud}-{\cal C}_{1du})\frac{c_3}{\sqrt{3}B_0F_\pi^2}(m_d-m_u)  \\
&&\!\!\!\!\times \frac{B_0F_\pi^2m_s+24a_0}{B_0F_\pi^2m_um_dm_s+8a_0(m_um_d+m_dm_s+m_sm_u)},
 \nn \\[1.5ex]
\label{eta0vev}
\frac{\langle\eta_0\rangle}{F_0}\simeq {} &&
\frac{G_F}{\sqrt{2}}({\cal C}_{1ud}-{\cal C}_{1du})\frac{\sqrt{2}c_3}{\sqrt{3}B_0F_\pi^2}(m_d-m_u)  \\
&&\!\!\!\!\times \frac{B_0F_\pi^2m_s}{B_0F_\pi^2m_um_dm_s+8a_0(m_um_d+m_dm_s+m_sm_u)} ,
\nn
\eea
where the leading short distance coefficients ${\cal C}_{1qq'}$ are given in \eq{C12ud-LR}.  The
comparison of \eqs{pivev}{eta0vev} shows the $m_s/(m_d-m_u)$ enhancement of $\vev{\pi^0}$ over the
others (empirically $20 a_0\simeq B_0 F_\pi^2 m_s$).

The relevant baryon chiral Lagrangian can be written as~\cite{Pich:1991fq}
\bea
{\cal L}_{\rm B}={}&&{\rm Tr}\left[\bar{B}i\gamma^\mu(\partial_\mu B+[\Gamma_\mu,\,B])-M_B \bar{B}B\right]
\nonumber \\[1.0ex]
&&-\frac{D}{2}{\rm tr}\left[\bar{B}\gamma^\mu\gamma_5\{\xi_\mu,\,B\}\right]
-\frac{F}{2}{\rm tr}\left[\bar{B}\gamma^\mu\gamma_5[\xi_\mu,\,B]\right]
\nonumber \\[1.0ex]
&&-\frac{\lambda}{2} \ {\rm tr}\left[\xi_\mu\right]{\rm tr}\left[\bar{B}\gamma^\mu\gamma_5B\right]
\nonumber \\[1.5ex]
&&+b_D \ {\rm tr}\left[\bar{B}\{\chi_+,\,B\}\right]
+b_F \ {\rm tr}\left[\bar{B}[\chi_+,\,B]\right]
\nonumber \\[2ex]
&&+b_0 \ {\rm tr}\left[\chi_+\right]{\rm tr}\left[\bar{B}B\right] + \cdots\,,
\label{Lbaryon}
\eea
 where
\begin{align}
&B=\begin{pmatrix}
      \frac{1}{\sqrt{2}}\Sigma^0+\frac{1}{\sqrt{6}}\Lambda^0 & \Sigma^+ & p \\
      \Sigma^- & -\frac{1}{\sqrt{2}}\Sigma^0+\frac{1}{\sqrt{6}}\Lambda^0 & n \\
      \Xi^- & \Xi^0 & -\frac{2}{\sqrt{6}}\Lambda^0 \\
   \end{pmatrix},\hspace{2em}
\\
&U=\xi_R\xi_L^\dagger    \qquad
(\xi_R=\xi_L^\dagger)
\end{align}
and
\begin{align}
 \Gamma_\mu\equiv {} &\frac{1}{2}\xi_R^\dagger(\partial_\mu-i\,r_\mu)\xi_R+\frac{1}{2}\xi_L^\dagger(\partial_\mu-i\,l_\mu)\xi_L\,,
 \\[1.5ex]
 \xi_\mu\equiv {} & i\xi_R^\dagger(\partial_\mu-i\,r_\mu)\xi_R-i\xi_L^\dagger(\partial_\mu-i\,l_\mu)\xi_L\,,
\\[2ex]
 \chi_+\equiv {} & \xi_L^\dagger \chi \xi_R+ \xi_R^\dagger \chi^\dagger \xi_L\,.
\end{align}
Finally, $M_B$ denotes the baryon mass in the chiral limit. In \eq{Lbaryon} the interaction terms
proportional to $D$, $F$ and $\lambda$ are CP conserving, while those proportional to $b_D$, $b_F$
and $b_0$ violate CP. The constants $D$ and $F$ are extracted from baryon semi-leptonic decays to be
at tree level $D\simeq 0.8$ and $F\simeq 0.5$~\cite{Scherer:2012xha}, while from baryon mass
splittings one obtains $b_D\simeq 0.07\,\rm GeV^{-1}$, 
$b_F\simeq -0.21\,\rm GeV^{-1}$~\cite{Pich:1991fq} and $b_0\simeq -0.52\,\rm GeV^{-1}$ from the $\pi N$ $\sigma$-term.

\begin{figure*}[t]
  \centerline{\includegraphics[width=0.95\columnwidth]{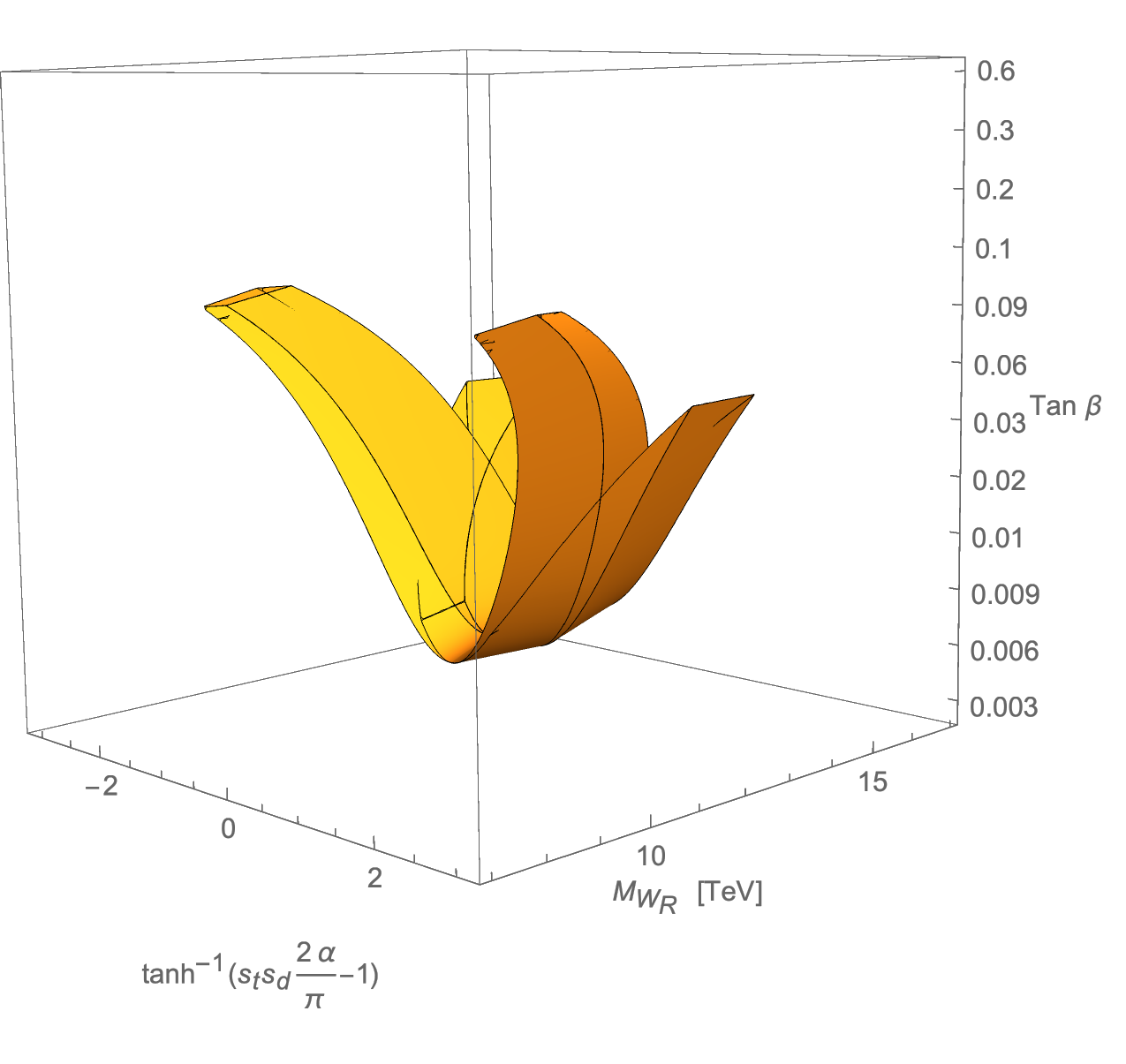}~~~%
    \vspace*{-3.5ex}
    \includegraphics[width=0.95\columnwidth]{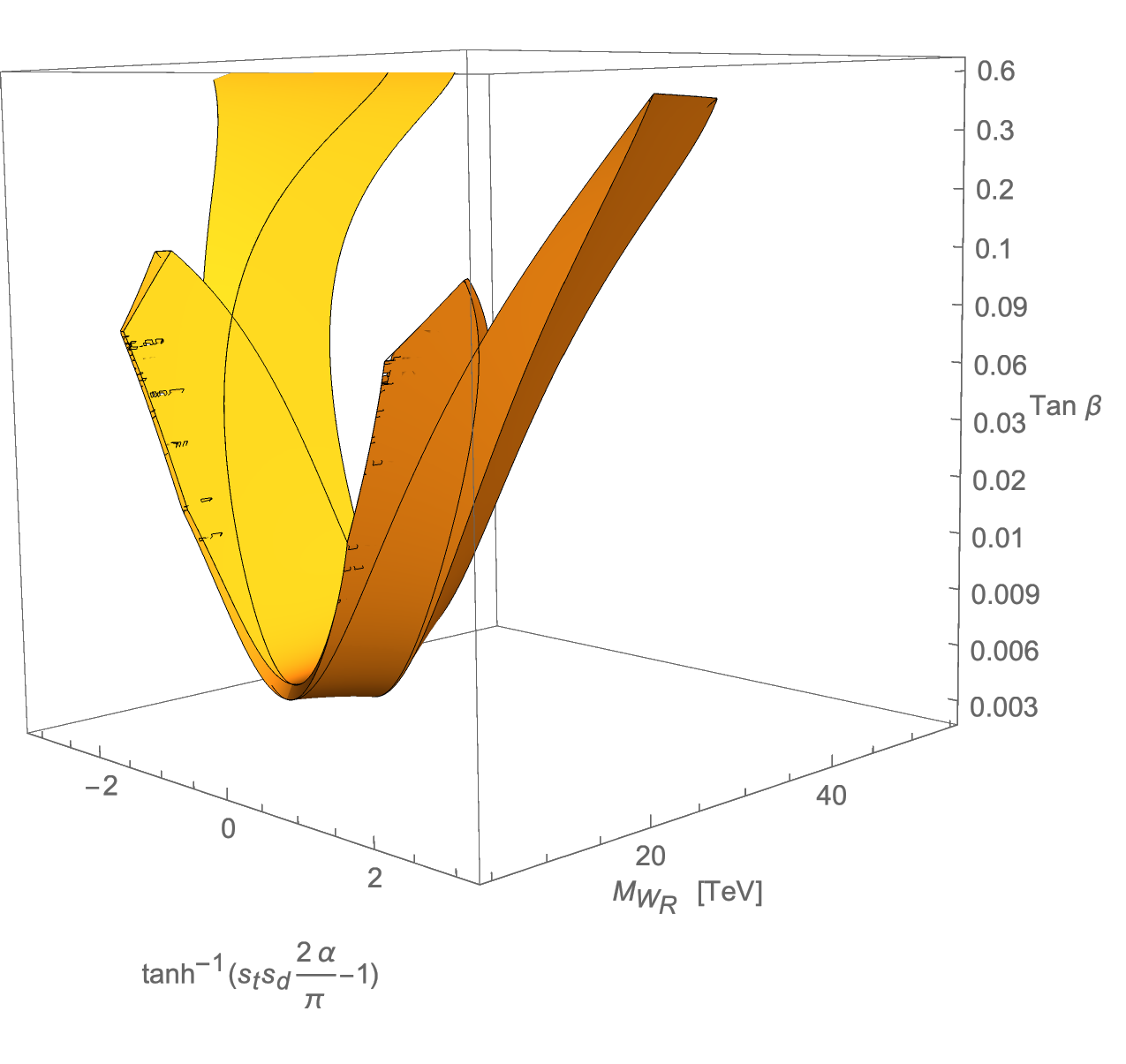}}
\caption{Case of $\P$: The allowed region in the $M_{W_R}$ -- $t_\b$ -- $\alpha$ space for the \NP scenario.
  \label{fig:P3D}}
    \vspace*{-2ex}
\end{figure*}

By properly rotating the meson fields in the baryonic lagrangian in such a way that $\vev{U}=1$ 
we can extract the relevant CP-violating baryon interactions with the physical meson fields~\cite{Pich:1991fq,An:2009zh}. 
Considering the vertices with one neutron and charged particles we obtain
\bea
\label{nppi}
\bar{g}_{np\pi}&=&\frac{B_0}{F_\pi}(b_D+b_F)\left[
2\sqrt{2}(m_d-m_u)\frac{\langle\pi^0\rangle}{F_\pi}\right.
  \\[0ex]
&&\left. -\frac{2\sqrt{2}}{\sqrt{3}}(m_u+m_d)\left(\frac{\langle\eta_8\rangle}{F_\pi}+\sqrt{2}\frac{\langle\eta_0\rangle}{F_0}\right)\right],
\nn\\[0.5ex]
\bar{g}_{n\Sigma^-K^+}&=&\frac{B_0}{F_\pi}(b_D-b_F)\left[
-2\sqrt{2}\ m_u\frac{\langle\pi^0\rangle}{F_\pi}\right.
  \\[0ex]
&&\left. -\frac{2\sqrt{2}}{\sqrt{3}}(m_u\!-\!2m_s)\frac{\langle\eta_8\rangle}{F_\pi}
-\frac{4}{\sqrt{3}}(m_u\!+\!m_s)\frac{\langle\eta_0\rangle}{F_0}\right] .  \nn 
\label{nsk}
\eea
These results differ from those obtained in Ref.~\cite{Haba:2018byj} by shifting linearly the meson
fields, a procedure that is bound to fail beyond the tadpole terms.  In particular, the coupling
$\bar{g}_{n\Sigma^-K^+}$ is no longer enhanced by $m_s/(m_d-m_u)$ over $\bar{g}_{np\pi}$, and while
the latter is dominated by $\langle\pi^0\rangle$ all VEVs contribute equally to the former. As a
consequence the logarithmic enhanced pion loops in \eq{dn_pi} still numerically dominate over the
kaon mediated contribution (\eq{dn_sigma}).

Similar considerations hold for $\bar{g}_{n\Sigma^0K^0}$ and $\bar{g}_{n\Lambda K}$, while for the
isovector coupling $\bar{g}_{nn\pi}$, relevant for the NLO pion loop contributions, we confirm the
result in~\cite{Haba:2018byj}:
\bea
&&\bar{g}_{nn\pi}=\frac{4B_0}{F_\pi}\Bigg[
\Big(\!\!-b_0(m_u\!+\!m_d)-(b_D\!+\!b_F)m_d\Big)\frac{\langle\pi^0\rangle}{F_\pi}
\nn\\
&&+\Big (b_0(m_d\!-\!m_u)+(b_D\!+\!b_F)m_d\Big)\left(\frac{\langle\eta_8\rangle}{\sqrt{3}F_\pi}
    +\sqrt{\frac23}\frac{\langle\eta_0\rangle}{F_0}\right)\Bigg]\,.\nn 
\label{nnp}\nn\\
\eea
%


When the LR scenario is endowed with a Peccei-Quinn symmetry the topological $\theta$-term can be
rotated away by an appropriate axion dependent chiral rotation of the quark fields
\beq
q_L\to e^{-i\,\a_q/2}q_L\,,\qquad  q_R\to e^{i\,\a_q/2}q_R\,,
\label{chirot}
\eeq
 where $\a_q$ depend on the axion field $a$ as
 \bea
\a_u&=&\frac{m_d m_s}{m_u m_d+m_d m_s+m_s m_u}\left(\frac{a}{f_a}+\bar{\theta}\right)
\nn \\[1ex]
\a_d&=&\frac{m_s m_u}{m_u m_d+m_d m_s+m_s m_u}\left(\frac{a}{f_a}+\bar{\theta}\right)
\nn \\[1ex]
\a_s&=&\frac{m_u m_d}{m_u m_d+m_d m_s+m_s m_u}\left(\frac{a}{f_a}+\bar{\theta}\right)
\label{chiphases}
\eea
and $f_a$ denotes the axion decay constant.  With the chosen $\a_q$ the axion does not mix with
$\pi^0$ and $\eta_8$.  By applying such an $U(3)_A$ field transformation to \eq{Lmeson}, the axion
field is included in the meson Lagrangian.

When only the leading $({\cal C}_{1ud}-{\cal C}_{1du})$ term is kept, the vacuum is readily obtained as
\bea
&&\frac{\langle\pi^0\rangle}{F_\pi}\simeq
\frac{G_F}{\sqrt{2}}({\cal C}_{1ud}-{\cal C}_{1du})\frac{c_3}{B_0F_\pi^2}
\frac{m_u+m_d+4m_s}{m_um_d+m_dm_s+m_sm_u}
\nn \\[1ex]
&&\frac{\langle\eta_8\rangle}{F_\pi}\simeq
\frac{G_F}{\sqrt{2}}({\cal C}_{1ud}-{\cal C}_{1du})\frac{\sqrt{3}c_3}{B_0F_\pi^2}
\frac{m_d-m_u}{m_um_d+m_dm_s+m_sm_u}
\nn \\[1ex]
&&\frac{\langle a\rangle}{f_a}+\bar{\theta}\simeq
\frac{G_F}{\sqrt{2}}({\cal C}_{1ud}-{\cal C}_{1du})\frac{2c_3}{B_0F_\pi^2}
\frac{m_d-m_u}{m_um_d}\,.
\label{PQvev}
\eea

The meson VEVs above follow closely
the pattern of \eqs{pivev}{eta8vev}, with $\vev{\eta_0}=0$, due to the dominance of the $a_0$ terms. Notice that the pion VEV is odd when exchanging $u$ and $d$, while the other VEVs are even as it must be.
Due to the presence of CP and chiral breaking effective LR operators, the axion VEV no longer
cancels the original $\bar{\theta}$ term, leaving a calculable $\bar{\theta}_{\rm eff}$
 (\eq{PQvev}) that contributes to the neutron EDM via an additional term in \eqs{nppi}{nsk}. We find
 \bea
&&\bar{g}_{np\pi}^{\, \theta} = - 4\sqrt{2}\frac{B_0}{F_\pi}(b_D+b_F) \frac{m_u m_d m_s\ \bar{\theta}_{\rm eff}}{m_sm_d+m_sm_u + m_u m_d}\ , \nn \\
&&\bar{g}_{n\Sigma^-K^+}^{\, \theta} = - 4\sqrt{2}\frac{B_0}{F_\pi}(b_D-b_F) \frac{m_u m_d m_s\ \bar{\theta}_{\rm eff}}{m_sm_d+m_sm_u + m_u m_d}\ ,
\label{PQaxvev} \nn \\[-0.5ex]
&&
\eea
which numerically lead to a contribution to the $n$EDM in good agreement with \eq{ThetaEDMn}.
It is noteworthy that by inserting the meson VEVs of \eq{PQvev} in \eq{PQaxvev} the $\bar{g}_{np\pi}$ vanishes identically.
On the other hand, when the ${\cal O}_1^{us}$ is considered the $\bar{g}_{n\Sigma^-K^+}$ coupling is in turn canceled, as consistency requires.
We have double checked this result using the formalism of Ref.~\cite{Pich:1991fq}, which makes such a cancelation more transparent.
 
In passing, let us note that also the axion mass is
modified with respect to the standard result by the presence of the new CP and chiral breaking
operators, but the deviation turns out to be utterly small.

\section{Numerical analysis for the case of $\P$}
\label{app:Pplots}

\noindent
For the case $\P$ there are 32 sign combinations $\{s_u,s_c,s_t,s_d,s_s\}$, corresponding to any
choice of them being $\pm 1$, after having conventionally set $s_b=1$.  They give rise to
different predictions for the $V_R$ phases $\theta_i$, as shown e.g.\ in
appendix~\ref{app:phases}. As a consequence, the numerical analysis has to be repeated separately for each
combination of signs.

For the \NP scenario, we find that one can accommodate simultaneously the $h_\eps$, $h_{\eps'}$ and
$h_{d_n}$ constraints only for $s_ss_d=1$; this is needed to avoid a $\pi$ shift which would lead to
a cancelation between the two terms in \eq{eq:epspsimple}. By inspection one finds also the
condition ${\rm sgn}\,\alpha=s_d s_t$, so that for numerical convenience one can restrict the
analysis to the ``log''-variable
$a=\tanh^{-1}\left[(2 s_d s_t \alpha/\pi) -1\right]\in(-\infty,\infty)$.  Finally, one finds that
solutions exist only in four cases: $\{1,1,1,1,1\}$, $\{1,1,1,-1,-1\}$, $\{-1,-1,-1,1,1\}$,
$\{-1,-1,-1,-1,-1\}$, the last two being just replicas of the first two.  The simultaneous
experimental constraints produce allowed regions in the $M_{W_R}$ -- $t_\b$ -- $a$ space, depicted
in \fig{fig:P3D}.  The numerical analysis is carried out for central values of the matrix elements
as well as for the enlarged conservative range, left and right frames of \fig{fig:P3D}
respectively. In this case solutions are found for four sign combinations more.  When the allowed
volumes are projected on the $M_{W_R}$--$t_\b$ plane, \fig{fig:masterPnp} is obtained.

A similar procedure is followed for the \SM scenario, where all 32 sign choices and both
${\rm sgn}\,\a=\pm 1$ contribute.  Here the lowest bound on the LR scale is found in the subset with
$s_cs_t=1$, where the $h_\eps$ constraint is easier to satisfy.




\bibliographystyle{utphysmod}
\bibliography{epsedm}

\providecommand{\href}[2]{#2}\begingroup\raggedright\begin{thebibliography}{100}

\bibitem{Glashow:1970gm}
S.~L. Glashow, J.~Iliopoulos, and L.~Maiani, ``{Weak Interactions with
  Lepton-Hadron Symmetry}'',
\href{http://dx.doi.org/10.1103/PhysRevD.2.1285}{\blue Phys. Rev. {\bfseries
  D2} (1970) 1285--1292}.

\bibitem{Kobayashi:1973fv}
M.~Kobayashi and T.~Maskawa, ``{CP Violation in the Renormalizable Theory of
  Weak Interaction}'',
\href{http://dx.doi.org/10.1143/PTP.49.652}{\blue Prog. Theor. Phys. {\bfseries
  49} (1973) 652--657}.

\bibitem{Engel:2013lsa}
J.~Engel, M.~J. Ramsey-Musolf, and U.~van Kolck, ``{Electric Dipole Moments of
  Nucleons, Nuclei, and Atoms: The Standard Model and Beyond}'',
  \href{http://dx.doi.org/10.1016/j.ppnp.2013.03.003}{\blue Prog. Part. Nucl.
  Phys. {\bfseries 71} (2013) 21--74},
\href{http://arxiv.org/abs/1303.2371}{{\blue arXiv:1303.2371 [nucl-th]}}.

\bibitem{Baker:2006ts}
C.~A. Baker {\em et~al.}, ``{An Improved experimental limit on the electric
  dipole moment of the neutron}'',
  \href{http://dx.doi.org/10.1103/PhysRevLett.97.131801}{\blue Phys. Rev. Lett.
  {\bfseries 97} (2006) 131801},
\href{http://arxiv.org/abs/hep-ex/0602020}{{\blue arXiv:hep-ex/0602020
  [hep-ex]}}.

\bibitem{Ellis:1978hq}
J.~R. Ellis and M.~K. Gaillard, ``{Strong and Weak CP Violation}'',
\href{http://dx.doi.org/10.1016/0550-3213(79)90297-9}{\blue Nucl. Phys.
  {\bfseries B150} (1979) 141--162}.

\bibitem{Pati:1974yy}
J.~C. Pati and A.~Salam, ``{Lepton Number as the Fourth Color}'',
  \href{http://dx.doi.org/10.1103/PhysRevD.10.275,
  10.1103/PhysRevD.11.703.2}{\blue Phys. Rev. {\bfseries D10} (1974) 275--289}.
[Erratum: Phys. Rev.D11,703(1975)].

\bibitem{Mohapatra:1974hk}
R.~N. Mohapatra and J.~C. Pati, ``{Left-Right Gauge Symmetry and an
  Isoconjugate Model of CP Violation}'',
\href{http://dx.doi.org/10.1103/PhysRevD.11.566}{\blue Phys. Rev. {\bfseries
  D11} (1975) 566--571}.

\bibitem{Mohapatra:1974gc}
R.~N. Mohapatra and J.~C. Pati, ``{A Natural Left-Right Symmetry}'',
\href{http://dx.doi.org/10.1103/PhysRevD.11.2558}{\blue Phys. Rev. {\bfseries
  D11} (1975) 2558}.

\bibitem{Senjanovic:1975rk}
G.~Senjanović and R.~N. Mohapatra, ``{Exact Left-Right Symmetry and
  Spontaneous Violation of Parity}'',
\href{http://dx.doi.org/10.1103/PhysRevD.12.1502}{\blue Phys. Rev. {\bfseries
  D12} (1975) 1502}.

\bibitem{Minkowski:1977sc}
P.~Minkowski, ``{$\mu \to e\gamma$ at a Rate of One Out of $10^{9}$ Muon
  Decays?}'',
\href{http://dx.doi.org/10.1016/0370-2693(77)90435-X}{\blue Phys. Lett.
  {\bfseries 67B} (1977) 421--428}.

\bibitem{Mohapatra:1979ia}
R.~N. Mohapatra and G.~Senjanović, ``{Neutrino Mass and Spontaneous Parity
  Nonconservation}'', \href{http://dx.doi.org/10.1103/PhysRevLett.44.912}{\blue
  Phys. Rev. Lett. {\bfseries 44} (1980) 912}.
[,231(1979)].

\bibitem{Senjanovic:1978ev}
G.~Senjanović, ``{Spontaneous Breakdown of Parity in a Class of Gauge
  Theories}'',
\href{http://dx.doi.org/10.1016/0550-3213(79)90604-7}{\blue Nucl. Phys.
  {\bfseries B153} (1979) 334--364}.

\bibitem{Mohapatra:1980yp}
R.~N. Mohapatra and G.~Senjanović, ``{Neutrino Masses and Mixings in Gauge
  Models with Spontaneous Parity Violation}'',
\href{http://dx.doi.org/10.1103/PhysRevD.23.165}{\blue Phys. Rev. {\bfseries
  D23} (1981) 165}.

\bibitem{Tello:2010am}
V.~Tello, M.~Nemevšek, F.~Nesti, G.~Senjanović, and F.~Vissani, ``{Left-Right
  Symmetry: from LHC to Neutrinoless Double Beta Decay}'',
  \href{http://dx.doi.org/10.1103/PhysRevLett.106.151801}{\blue Phys. Rev.
  Lett. {\bfseries 106} (2011) 151801},
\href{http://arxiv.org/abs/1011.3522}{{\blue arXiv:1011.3522 [hep-ph]}}.

\bibitem{Keung:1983uu}
W.-Y. Keung and G.~Senjanović, ``{Majorana Neutrinos and the Production of the
  Right-handed Charged Gauge Boson}'',
\href{http://dx.doi.org/10.1103/PhysRevLett.50.1427}{\blue Phys. Rev. Lett.
  {\bfseries 50} (1983) 1427}.

\bibitem{Nemevsek:2012iq}
M.~Nemevšek, G.~Senjanović, and V.~Tello, ``{Connecting Dirac and Majorana
  Neutrino Mass Matrices in the Minimal Left-Right Symmetric Model}'',
  \href{http://dx.doi.org/10.1103/PhysRevLett.110.151802}{\blue Phys. Rev.
  Lett. {\bfseries 110} no.~15, (2013) 151802},
\href{http://arxiv.org/abs/1211.2837}{{\blue arXiv:1211.2837 [hep-ph]}}.

\bibitem{Senjanovic:2016vxw}
G.~Senjanović and V.~Tello, ``{Probing Seesaw with Parity Restoration}'',
  \href{http://dx.doi.org/10.1103/PhysRevLett.119.201803}{\blue Phys. Rev.
  Lett. {\bfseries 119} no.~20, (2017) 201803},
\href{http://arxiv.org/abs/1612.05503}{{\blue arXiv:1612.05503 [hep-ph]}}.

\bibitem{Senjanovic:2018xtu}
G.~Senjanović and V.~Tello, ``{Disentangling the seesaw mechanism in the
  minimal left-right symmetric model}'',
  \href{http://dx.doi.org/10.1103/PhysRevD.100.115031}{\blue Phys. Rev.
  {\bfseries D100} no.~11, (2019) 115031},
\href{http://arxiv.org/abs/1812.03790}{{\blue arXiv:1812.03790 [hep-ph]}}.

\bibitem{Helo:2018rll}
J.~C. Helo, H.~Li, N.~A. Neill, M.~Ramsey-Musolf, and J.~C. Vasquez, ``{Probing
  neutrino Dirac mass in left-right symmetric models at the LHC and next
  generation colliders}'',
  \href{http://dx.doi.org/10.1103/PhysRevD.99.055042}{\blue Phys. Rev.
  {\bfseries D99} no.~5, (2019) 055042},
\href{http://arxiv.org/abs/1812.01630}{{\blue arXiv:1812.01630 [hep-ph]}}.

\bibitem{Beall:1981ze}
G.~Beall, M.~Bander, and A.~Soni, ``{Constraint on the Mass Scale of a
  Left-Right Symmetric Electroweak Theory from the K(L) K(S) Mass
  Difference}'',
\href{http://dx.doi.org/10.1103/PhysRevLett.48.848}{\blue Phys. Rev. Lett.
  {\bfseries 48} (1982) 848}.

\bibitem{Ecker:1985vv}
G.~Ecker and W.~Grimus, ``{CP Violation and Left-Right Symmetry}'',
\href{http://dx.doi.org/10.1016/0550-3213(85)90616-9}{\blue Nucl. Phys.
  {\bfseries B258} (1985) 328--360}.

\bibitem{Maiezza:2010ic}
A.~Maiezza, M.~Nemevšek, F.~Nesti, and G.~Senjanović, ``{Left-Right Symmetry
  at LHC}'', \href{http://dx.doi.org/10.1103/PhysRevD.82.055022}{\blue Phys.
  Rev. {\bfseries D82} (2010) 055022},
\href{http://arxiv.org/abs/1005.5160}{{\blue arXiv:1005.5160 [hep-ph]}}.

\bibitem{Bertolini:2012pu}
S.~Bertolini, J.~O. Eeg, A.~Maiezza, and F.~Nesti, ``{New physics in
  $\epsilon'$ from gluomagnetic contributions and limits on Left-Right
  symmetry}'', \href{http://dx.doi.org/10.1103/PhysRevD.86.095013,
  10.1103/PhysRevD.93.079903}{\blue Phys. Rev. {\bfseries D86} (2012) 095013},
  \href{http://arxiv.org/abs/1206.0668}{{\blue arXiv:1206.0668 [hep-ph]}}.
[Erratum: Phys. Rev.D93,no.7,079903(2016)].

\bibitem{Bertolini:2013noa}
S.~Bertolini, A.~Maiezza, and F.~Nesti, ``{$K\to\pi\pi$ hadronic matrix
  elements of left-right current-current operators}'',
  \href{http://dx.doi.org/10.1103/PhysRevD.88.034014}{\blue Phys. Rev.
  {\bfseries D88} no.~3, (2013) 034014},
\href{http://arxiv.org/abs/1305.5739}{{\blue arXiv:1305.5739 [hep-ph]}}.

\bibitem{Bertolini:2014sua}
S.~Bertolini, A.~Maiezza, and F.~Nesti, ``{Present and Future K and B Meson
  Mixing Constraints on TeV Scale Left-Right Symmetry}'',
  \href{http://dx.doi.org/10.1103/PhysRevD.89.095028}{\blue Phys. Rev.
  {\bfseries D89} no.~9, (2014) 095028},
\href{http://arxiv.org/abs/1403.7112}{{\blue arXiv:1403.7112 [hep-ph]}}.

\bibitem{Mohapatra:1978fy}
R.~N. Mohapatra and G.~Senjanović, ``{Natural Suppression of Strong p and t
  Noninvariance}'',
\href{http://dx.doi.org/10.1016/0370-2693(78)90243-5}{\blue Phys. Lett.
  {\bfseries 79B} (1978) 283--286}.

\bibitem{Peccei:1977hh}
R.~D. Peccei and H.~R. Quinn, ``{CP Conservation in the Presence of
  Instantons}'', \href{http://dx.doi.org/10.1103/PhysRevLett.38.1440}{\blue
  Phys. Rev. Lett. {\bfseries 38} (1977) 1440--1443}.
[,328(1977)].

\bibitem{Weinberg:1977ma}
S.~Weinberg, ``{A New Light Boson?}'',
\href{http://dx.doi.org/10.1103/PhysRevLett.40.223}{\blue Phys. Rev. Lett.
  {\bfseries 40} (1978) 223--226}.

\bibitem{Wilczek:1977pj}
F.~Wilczek, ``{Problem of Strong $P$ and $T$ Invariance in the Presence of
  Instantons}'',
\href{http://dx.doi.org/10.1103/PhysRevLett.40.279}{\blue Phys. Rev. Lett.
  {\bfseries 40} (1978) 279--282}.

\bibitem{Maiezza:2014ala}
A.~Maiezza and M.~Nemevšek, ``{Strong P invariance, neutron electric dipole
  moment, and minimal left-right parity at LHC}'',
  \href{http://dx.doi.org/10.1103/PhysRevD.90.095002}{\blue Phys. Rev.
  {\bfseries D90} no.~9, (2014) 095002},
\href{http://arxiv.org/abs/1407.3678}{{\blue arXiv:1407.3678 [hep-ph]}}.

\bibitem{Maiezza:2016bzp}
A.~Maiezza, M.~Nemevšek, and F.~Nesti, ``{Perturbativity and mass scales in
  the minimal left-right symmetric model}'',
  \href{http://dx.doi.org/10.1103/PhysRevD.94.035008}{\blue Phys. Rev.
  {\bfseries D94} no.~3, (2016) 035008},
\href{http://arxiv.org/abs/1603.00360}{{\blue arXiv:1603.00360 [hep-ph]}}.

\bibitem{Maiezza:2016ybz}
A.~Maiezza, G.~Senjanović, and J.~C. Vasquez, ``{Higgs sector of the minimal
  left-right symmetric theory}'',
  \href{http://dx.doi.org/10.1103/PhysRevD.95.095004}{\blue Phys. Rev.
  {\bfseries D95} no.~9, (2017) 095004},
\href{http://arxiv.org/abs/1612.09146}{{\blue arXiv:1612.09146 [hep-ph]}}.

\bibitem{Chauhan:2018uuy}
G.~Chauhan, P.~S.~B. Dev, R.~N. Mohapatra, and Y.~Zhang, ``{Perturbativity
  constraints on $U(1)_{B-L}$ and left-right models and implications for heavy
  gauge boson searches}'',
  \href{http://dx.doi.org/10.1007/JHEP01(2019)208}{\blue JHEP {\bfseries 01}
  (2019) 208},
\href{http://arxiv.org/abs/1811.08789}{{\blue arXiv:1811.08789 [hep-ph]}}.

\bibitem{Nemevsek:2018bbt}
M.~Nemevšek, F.~Nesti, and G.~Popara, ``{Keung-Senjanović process at the LHC:
  From lepton number violation to displaced vertices to invisible decays}'',
  \href{http://dx.doi.org/10.1103/PhysRevD.97.115018}{\blue Phys. Rev.
  {\bfseries D97} no.~11, (2018) 115018},
\href{http://arxiv.org/abs/1801.05813}{{\blue arXiv:1801.05813 [hep-ph]}}.

\bibitem{Gisbert:2017vvj}
H.~Gisbert and A.~Pich, ``{Direct CP violation in $K^0\to\pi\pi$: Standard
  Model Status}'', \href{http://dx.doi.org/10.1088/1361-6633/aac18e}{\blue
  Rept. Prog. Phys. {\bfseries 81} no.~7, (2018) 076201},
\href{http://arxiv.org/abs/1712.06147}{{\blue arXiv:1712.06147 [hep-ph]}}.

\bibitem{Buras:2018ozh}
A.~J. Buras, ``{$\epsilon^\prime/\epsilon$-2018: A Christmas Story}'',
\href{http://arxiv.org/abs/1812.06102}{{\blue arXiv:1812.06102 [hep-ph]}}.

\bibitem{Frere:1990cj}
J.~M. Frere, J.~Galand, A.~Le~Yaouanc, L.~Oliver, O.~Pene, and J.~C. Raynal,
  ``{QCD corrections to the quark electric dipole moment in the SU(2)-L x
  SU(2)-R x U(1) model of CP violation}'',
\href{http://dx.doi.org/10.1016/0370-2693(90)90733-M}{\blue Phys. Lett.
  {\bfseries B251} (1990) 443--449}.

\bibitem{Frere:1991jt}
J.~M. Frere, J.~Galand, A.~Le~Yaouanc, L.~Oliver, O.~Pene, and J.~C. Raynal,
  ``{The Neutron electric dipole moment in left-right symmetric models}'',
\href{http://dx.doi.org/10.1103/PhysRevD.45.259}{\blue Phys. Rev. {\bfseries
  D45} (1992) 259--277}.

\bibitem{Frere:1991db}
J.~M. Frere, J.~Galand, A.~Le~Yaouanc, L.~Oliver, O.~Pene, and J.~C. Raynal,
  ``{K0 anti-K0 in the SU(2)-L x SU(2)-R x U(1) model of CP violation}'',
\href{http://dx.doi.org/10.1103/PhysRevD.46.337}{\blue Phys. Rev. {\bfseries
  D46} (1992) 337--353}.

\bibitem{Cirigliano:2016yhc}
V.~Cirigliano, W.~Dekens, J.~de~Vries, and E.~Mereghetti, ``{An $\epsilon'$
  improvement from right-handed currents}'',
  \href{http://dx.doi.org/10.1016/j.physletb.2017.01.037}{\blue Phys. Lett.
  {\bfseries B767} (2017) 1--9},
\href{http://arxiv.org/abs/1612.03914}{{\blue arXiv:1612.03914 [hep-ph]}}.

\bibitem{Dekens:2017hyc}
W.~Dekens, ``{$\epsilon'$ from right-handed currents}'', in {\em {Proceedings,
  52nd Rencontres de Moriond on Electroweak Interactions and Unified Theories:
  La Thuile, Italy, March 18-25, 2017}}, pp.~187--194.
\newblock 2017.
\newblock
\href{http://arxiv.org/abs/1708.00797}{{\blue arXiv:1708.00797 [hep-ph]}}.
\newblock

\bibitem{Haba:2018byj}
N.~Haba, H.~Umeeda, and T.~Yamada, ``{$\epsilon'/\epsilon$ Anomaly and Neutron
  EDM in $SU(2)_L\times SU(2)_R\times U(1)_{B-L}$ model with Charge
  Symmetry}'', \href{http://dx.doi.org/10.1007/JHEP05(2018)052}{\blue JHEP
  {\bfseries 05} (2018) 052},
\href{http://arxiv.org/abs/1802.09903}{{\blue arXiv:1802.09903 [hep-ph]}}.

\bibitem{Haba:2017jgf}
N.~Haba, H.~Umeeda, and T.~Yamada, ``{Semialigned two Higgs doublet model}'',
  \href{http://dx.doi.org/10.1103/PhysRevD.97.035004}{\blue Phys. Rev.
  {\bfseries D97} no.~3, (2018) 035004},
\href{http://arxiv.org/abs/1711.06499}{{\blue arXiv:1711.06499 [hep-ph]}}.

\bibitem{Senjanovic:2014pva}
G.~Senjanović and V.~Tello, ``{Right Handed Quark Mixing in Left-Right
  Symmetric Theory}'',
  \href{http://dx.doi.org/10.1103/PhysRevLett.114.071801}{\blue Phys. Rev.
  Lett. {\bfseries 114} no.~7, (2015) 071801},
\href{http://arxiv.org/abs/1408.3835}{{\blue arXiv:1408.3835 [hep-ph]}}.

\bibitem{Senjanovic:2015yea}
G.~Senjanović and V.~Tello, ``{Restoration of Parity and the Right-Handed
  Analog of the CKM Matrix}'',
  \href{http://dx.doi.org/10.1103/PhysRevD.94.095023}{\blue Phys. Rev.
  {\bfseries D94} no.~9, (2016) 095023},
\href{http://arxiv.org/abs/1502.05704}{{\blue arXiv:1502.05704 [hep-ph]}}.

\bibitem{Senjanovic:1979cta}
G.~Senjanović and P.~Senjanović, ``{Suppression of Higgs Strangeness Changing
  Neutral Currents in a Class of Gauge Theories}'',
\href{http://dx.doi.org/10.1103/PhysRevD.21.3253}{\blue Phys. Rev. {\bfseries
  D21} (1980) 3253}.

\bibitem{Basecq:1985cr}
J.~Basecq, L.-F. Li, and P.~B. Pal, ``{Gauge Invariant Calculation of the $K_L
  K_S$ Mass Difference in the Left-right Model}'',
\href{http://dx.doi.org/10.1103/PhysRevD.32.175}{\blue Phys. Rev. {\bfseries
  D32} (1985) 175}.

\bibitem{Aoki:2019cca}
{ Flavour Lattice Averaging Group}, S.~Aoki {\em et~al.}, ``{FLAG Review
  2019}'',
\href{http://arxiv.org/abs/1902.08191}{{\blue arXiv:1902.08191 [hep-lat]}}.

\bibitem{Buras:2013ooa}
A.~J. Buras and J.~Girrbach, ``{Towards the Identification of New Physics
  through Quark Flavour Violating Processes}'',
  \href{http://dx.doi.org/10.1088/0034-4885/77/8/086201}{\blue Rept. Prog.
  Phys. {\bfseries 77} (2014) 086201},
\href{http://arxiv.org/abs/1306.3775}{{\blue arXiv:1306.3775 [hep-ph]}}.

\bibitem{Gaillard:1974hs}
M.~K. Gaillard and B.~W. Lee, ``{Rare Decay Modes of the K-Mesons in Gauge
  Theories}'',
\href{http://dx.doi.org/10.1103/PhysRevD.10.897}{\blue Phys. Rev. {\bfseries
  D10} (1974) 897}.

\bibitem{Manohar:1983md}
A.~Manohar and H.~Georgi, ``{Chiral Quarks and the Nonrelativistic Quark
  Model}'',
\href{http://dx.doi.org/10.1016/0550-3213(84)90231-1}{\blue Nucl. Phys.
  {\bfseries B234} (1984) 189--212}.

\bibitem{Cohen:1984vv}
A.~G. Cohen and A.~V. Manohar, ``{The $\Delta I = 1/2$ Rule in the Chiral Quark
  Model}'',
\href{http://dx.doi.org/10.1016/0370-2693(84)91506-5}{\blue Phys. Lett.
  {\bfseries 143B} (1984) 481--484}.

\bibitem{Bertolini:1998vd}
S.~Bertolini, M.~Fabbrichesi, and J.~O. Eeg, ``{Theory of the CP violating
  parameter epsilon-prime / epsilon}'',
  \href{http://dx.doi.org/10.1103/RevModPhys.72.65}{\blue Rev. Mod. Phys.
  {\bfseries 72} (2000) 65--93},
\href{http://arxiv.org/abs/hep-ph/9802405}{{\blue arXiv:hep-ph/9802405
  [hep-ph]}}.

\bibitem{Bertolini:1997ir}
S.~Bertolini, J.~O. Eeg, M.~Fabbrichesi, and E.~I. Lashin, ``{The Delta I = 1/2
  rule and B(K) at O (p**4) in the chiral expansion}'',
  \href{http://dx.doi.org/10.1016/S0550-3213(97)00787-6}{\blue Nucl. Phys.
  {\bfseries B514} (1998) 63--92},
\href{http://arxiv.org/abs/hep-ph/9705244}{{\blue arXiv:hep-ph/9705244
  [hep-ph]}}.

\bibitem{Antonelli:1995nv}
V.~Antonelli, S.~Bertolini, J.~O. Eeg, M.~Fabbrichesi, and E.~I. Lashin, ``{The
  Delta S = 1 weak chiral lagrangian as the effective theory of the chiral
  quark model}'', \href{http://dx.doi.org/10.1016/0550-3213(96)00144-7}{\blue
  Nucl. Phys. {\bfseries B469} (1996) 143--180},
\href{http://arxiv.org/abs/hep-ph/9511255}{{\blue arXiv:hep-ph/9511255
  [hep-ph]}}.

\bibitem{Bertolini:1993rc}
S.~Bertolini, M.~Fabbrichesi, and E.~Gabrielli, ``{The Relevance of the dipole
  Penguin operators in epsilon-prime / epsilon}'',
  \href{http://dx.doi.org/10.1016/0370-2693(94)91541-5}{\blue Phys. Lett.
  {\bfseries B327} (1994) 136--144},
\href{http://arxiv.org/abs/hep-ph/9312266}{{\blue arXiv:hep-ph/9312266
  [hep-ph]}}.

\bibitem{Bertolini:1994qk}
S.~Bertolini, J.~O. Eeg, and M.~Fabbrichesi, ``{Studying epsilon-prime /
  epsilon in the chiral quark model: gamma(5) scheme independence and NLO
  hadronic matrix elements}'',
  \href{http://dx.doi.org/10.1016/0550-3213(95)00274-V}{\blue Nucl. Phys.
  {\bfseries B449} (1995) 197--228},
\href{http://arxiv.org/abs/hep-ph/9409437}{{\blue arXiv:hep-ph/9409437
  [hep-ph]}}.

\bibitem{Constantinou:2017sgv}
{ ETM}, M.~Constantinou, M.~Costa, R.~Frezzotti, V.~Lubicz, G.~Martinelli,
  D.~Meloni, H.~Panagopoulos, and S.~Simula, ``{$K \to \pi$ matrix elements of
  the chromomagnetic operator on the lattice}'',
  \href{http://dx.doi.org/10.1103/PhysRevD.97.074501}{\blue Phys. Rev.
  {\bfseries D97} no.~7, (2018) 074501},
\href{http://arxiv.org/abs/1712.09824}{{\blue arXiv:1712.09824 [hep-lat]}}.

\bibitem{Buras:2018evv}
A.~J. Buras and J.-M. G\'erard, ``{$K\to\pi\pi$ and $K-\pi$ Matrix Elements of
  the Chromomagnetic Operators from Dual QCD}'',
  \href{http://dx.doi.org/10.1007/JHEP07(2018)126}{\blue JHEP {\bfseries 07}
  (2018) 126},
\href{http://arxiv.org/abs/1803.08052}{{\blue arXiv:1803.08052 [hep-ph]}}.

\bibitem{Bertolini:1995tp}
S.~Bertolini, J.~O. Eeg, and M.~Fabbrichesi, ``{A New estimate of epsilon-prime
  / epsilon}'', \href{http://dx.doi.org/10.1016/0550-3213(96)00335-5}{\blue
  Nucl. Phys. {\bfseries B476} (1996) 225--254},
\href{http://arxiv.org/abs/hep-ph/9512356}{{\blue arXiv:hep-ph/9512356
  [hep-ph]}}.

\bibitem{Bertolini:1997nf}
S.~Bertolini, J.~O. Eeg, M.~Fabbrichesi, and E.~I. Lashin, ``{Epsilon-prime /
  epsilon at O(p**4) in the chiral expansion}'',
  \href{http://dx.doi.org/10.1016/S0550-3213(97)00786-4}{\blue Nucl. Phys.
  {\bfseries B514} (1998) 93--112},
\href{http://arxiv.org/abs/hep-ph/9706260}{{\blue arXiv:hep-ph/9706260
  [hep-ph]}}.

\bibitem{AlaviHarati:1999xp}
{ KTeV}, A.~Alavi-Harati {\em et~al.}, ``{Observation of direct CP violation in
  $K_{S,L} \to \pi \pi$ decays}'',
  \href{http://dx.doi.org/10.1103/PhysRevLett.83.22}{\blue Phys. Rev. Lett.
  {\bfseries 83} (1999) 22--27},
\href{http://arxiv.org/abs/hep-ex/9905060}{{\blue arXiv:hep-ex/9905060
  [hep-ex]}}.

\bibitem{Fanti:1999nm}
{ NA48}, V.~Fanti {\em et~al.}, ``{A New measurement of direct CP violation in
  two pion decays of the neutral kaon}'',
  \href{http://dx.doi.org/10.1016/S0370-2693(99)01030-8}{\blue Phys. Lett.
  {\bfseries B465} (1999) 335--348},
\href{http://arxiv.org/abs/hep-ex/9909022}{{\blue arXiv:hep-ex/9909022
  [hep-ex]}}.

\bibitem{Pallante:1999qf}
E.~Pallante and A.~Pich, ``{Strong enhancement of epsilon-prime / epsilon
  through final state interactions}'',
  \href{http://dx.doi.org/10.1103/PhysRevLett.84.2568}{\blue Phys. Rev. Lett.
  {\bfseries 84} (2000) 2568--2571},
\href{http://arxiv.org/abs/hep-ph/9911233}{{\blue arXiv:hep-ph/9911233
  [hep-ph]}}.

\bibitem{Pallante:2000hk}
E.~Pallante and A.~Pich, ``{Final state interactions in kaon decays}'',
  \href{http://dx.doi.org/10.1016/S0550-3213(00)00601-5}{\blue Nucl. Phys.
  {\bfseries B592} (2001) 294--320},
\href{http://arxiv.org/abs/hep-ph/0007208}{{\blue arXiv:hep-ph/0007208
  [hep-ph]}}.

\bibitem{Pallante:2000pz}
E.~Pallante, A.~Pich, and I.~Scimemi, ``{The Role of final state interactions
  in epsilon-prime / epsilon}'',
  \href{http://dx.doi.org/10.1142/S0217751X01007765}{\blue Int. J. Mod. Phys.
  {\bfseries A16} (2001) 672--674},
\href{http://arxiv.org/abs/hep-ph/0010229}{{\blue arXiv:hep-ph/0010229
  [hep-ph]}}.

\bibitem{Buchler:2001nm}
M.~Buchler, G.~Colangelo, J.~Kambor, and F.~Orellana, ``{Dispersion relations
  and soft pion theorems for K to pi pi}'',
  \href{http://dx.doi.org/10.1016/S0370-2693(01)01098-X}{\blue Phys. Lett.
  {\bfseries B521} (2001) 22--28},
\href{http://arxiv.org/abs/hep-ph/0102287}{{\blue arXiv:hep-ph/0102287
  [hep-ph]}}.

\bibitem{Buchler:2001np}
M.~Buchler, G.~Colangelo, J.~Kambor, and F.~Orellana, ``{A Note on the
  dispersive treatment of K to pi pi with the kaon off-shell}'',
  \href{http://dx.doi.org/10.1016/S0370-2693(01)01197-2}{\blue Phys. Lett.
  {\bfseries B521} (2001) 29--32},
\href{http://arxiv.org/abs/hep-ph/0102289}{{\blue arXiv:hep-ph/0102289
  [hep-ph]}}.

\bibitem{Feinberg:1959ui}
G.~Feinberg, P.~Kabir, and S.~Weinberg, ``{Transformation of muons into
  electrons}'',
\href{http://dx.doi.org/10.1103/PhysRevLett.3.527}{\blue Phys. Rev. Lett.
  {\bfseries 3} (1959) 527--530}.

\bibitem{Buras:2016fys}
A.~J. Buras and J.-M. G\'erard, ``{Final state interactions in $K\rightarrow
  \pi \pi $ decays: $\Delta I=1/2$ rule vs. $\varepsilon '/\varepsilon $}'',
  \href{http://dx.doi.org/10.1140/epjc/s10052-016-4586-7}{\blue Eur. Phys. J.
  {\bfseries C77} no.~1, (2017) 10},
\href{http://arxiv.org/abs/1603.05686}{{\blue arXiv:1603.05686 [hep-ph]}}.

\bibitem{Aebischer:2018rrz}
J.~Aebischer, A.~J. Buras, and J.-M. G\'erard, ``{BSM hadronic matrix elements
  for $\epsilon'/\epsilon$ and $K\to\pi\pi$ decays in the Dual QCD approach}'',
  \href{http://dx.doi.org/10.1007/JHEP02(2019)021}{\blue JHEP {\bfseries 02}
  (2019) 021},
\href{http://arxiv.org/abs/1807.01709}{{\blue arXiv:1807.01709 [hep-ph]}}.

\bibitem{Bardeen:1986vp}
W.~A. Bardeen, A.~J. Buras, and J.~M. G\'erard, ``{The Delta I = 1/2 Rule in
  the Large N Limit}'',
\href{http://dx.doi.org/10.1016/0370-2693(86)90150-4}{\blue Phys. Lett.
  {\bfseries B180} (1986) 133--140}.

\bibitem{Bardeen:1986uz}
W.~A. Bardeen, A.~J. Buras, and J.~M. G\'erard, ``{The K $\to$ pi pi Decays in
  the Large n Limit: Quark Evolution}'',
\href{http://dx.doi.org/10.1016/0550-3213(87)90091-5}{\blue Nucl. Phys.
  {\bfseries B293} (1987) 787--811}.

\bibitem{Bardeen:1986vz}
W.~A. Bardeen, A.~J. Buras, and J.~M. G\'erard, ``{A Consistent Analysis of the
  Delta I = 1/2 Rule for K Decays}'',
\href{http://dx.doi.org/10.1016/0370-2693(87)91156-7}{\blue Phys. Lett.
  {\bfseries B192} (1987) 138--144}.

\bibitem{Fatelo:1994qh}
J.~P. Fatelo and J.~M. G\'erard, ``{Current current operator evolution in the
  chiral limit}'',
\href{http://dx.doi.org/10.1016/0370-2693(95)00047-O}{\blue Phys. Lett.
  {\bfseries B347} (1995) 136--142}.

\bibitem{Bardeen:1987vg}
W.~A. Bardeen, A.~J. Buras, and J.~M. G\'erard, ``{The B Parameter Beyond the
  Leading Order of 1/n Expansion}'',
\href{http://dx.doi.org/10.1016/0370-2693(88)90913-6}{\blue Phys. Lett.
  {\bfseries B211} (1988) 343--349}.

\bibitem{Buras:2014maa}
A.~J. Buras, J.-M. G\'erard, and W.~A. Bardeen, ``{Large $N$ Approach to Kaon
  Decays and Mixing 28 Years Later: $\Delta I = 1/2$ Rule, $\hat B_K$ and
  $\Delta M_K$}'',
  \href{http://dx.doi.org/10.1140/epjc/s10052-014-2871-x}{\blue Eur. Phys. J.
  {\bfseries C74} (2014) 2871},
\href{http://arxiv.org/abs/1401.1385}{{\blue arXiv:1401.1385 [hep-ph]}}.

\bibitem{Garron:2016mva}
{ RBC/UKQCD}, N.~Garron, R.~J. Hudspith, and A.~T. Lytle, ``{Neutral Kaon
  Mixing Beyond the Standard Model with $n_f=2+1$ Chiral Fermions Part 1: Bare
  Matrix Elements and Physical Results}'',
  \href{http://dx.doi.org/10.1007/JHEP11(2016)001}{\blue JHEP {\bfseries 11}
  (2016) 001},
\href{http://arxiv.org/abs/1609.03334}{{\blue arXiv:1609.03334 [hep-lat]}}.

\bibitem{Boyle:2017skn}
{ RBC, UKQCD}, P.~A. Boyle, N.~Garron, R.~J. Hudspith, C.~Lehner, and A.~T.
  Lytle, ``{Neutral kaon mixing beyond the Standard Model with n$_{f}$ = 2 + 1
  chiral fermions. Part 2: non perturbative renormalisation of the $\Delta F=2$
  four-quark operators}'',
  \href{http://dx.doi.org/10.1007/JHEP10(2017)054}{\blue JHEP {\bfseries 10}
  (2017) 054},
\href{http://arxiv.org/abs/1708.03552}{{\blue arXiv:1708.03552 [hep-lat]}}.

\bibitem{Buras:2018lgu}
A.~J. Buras and J.-M. G\'erard, ``{Dual QCD Insight into BSM Hadronic Matrix
  Elements for $K^0-\bar K^0$ Mixing from Lattice QCD}'',
  \href{http://dx.doi.org/10.5506/APhysPolB.50.121}{\blue Acta Phys. Polon.
  {\bfseries B50} (2019) 121},
\href{http://arxiv.org/abs/1804.02401}{{\blue arXiv:1804.02401 [hep-ph]}}.

\bibitem{Blum:2015ywa}
T.~Blum {\em et~al.}, ``{$K \rightarrow \pi\pi$ $\Delta I=3/2$ decay amplitude
  in the continuum limit}'',
  \href{http://dx.doi.org/10.1103/PhysRevD.91.074502}{\blue Phys. Rev.
  {\bfseries D91} no.~7, (2015) 074502},
\href{http://arxiv.org/abs/1502.00263}{{\blue arXiv:1502.00263 [hep-lat]}}.

\bibitem{Bai:2015nea}
{ RBC, UKQCD}, Z.~Bai {\em et~al.}, ``{Standard Model Prediction for Direct CP
  Violation in $K \rightarrow \pi\pi$ Decay}'',
  \href{http://dx.doi.org/10.1103/PhysRevLett.115.212001}{\blue Phys. Rev.
  Lett. {\bfseries 115} no.~21, (2015) 212001},
\href{http://arxiv.org/abs/1505.07863}{{\blue arXiv:1505.07863 [hep-lat]}}.

\bibitem{Aebischer:2019mtr}
J.~Aebischer, C.~Bobeth, and A.~J. Buras, ``{On the importance of NNLO QCD and
  isospin-breaking corrections in $\varepsilon '/\varepsilon $}'',
  \href{http://dx.doi.org/10.1140/epjc/s10052-019-7549-y}{\blue Eur. Phys. J.
  {\bfseries C80} no.~1, (2020) 1},
\href{http://arxiv.org/abs/1909.05610}{{\blue arXiv:1909.05610 [hep-ph]}}.

\bibitem{Cirigliano:2019cpi}
V.~Cirigliano, H.~Gisbert, A.~Pich, and A.~Rodriguez-Sanchez,
  ``{Isospin-Violating Contributions to $\epsilon'/\epsilon$}'',
\href{http://arxiv.org/abs/1911.01359}{{\blue arXiv:1911.01359 [hep-ph]}}.

\bibitem{Barr:1991qx}
S.~M. Barr, D.~Chang, and G.~Senjanović, ``{Strong CP problem and parity}'',
\href{http://dx.doi.org/10.1103/PhysRevLett.67.2765}{\blue Phys. Rev. Lett.
  {\bfseries 67} (1991) 2765--2768}.

\bibitem{Kuchimanchi:2010xs}
R.~Kuchimanchi, ``{P/CP Conserving CP/P Violation Solves Strong CP Problem}'',
  \href{http://dx.doi.org/10.1103/PhysRevD.82.116008}{\blue Phys. Rev.
  {\bfseries D82} (2010) 116008},
\href{http://arxiv.org/abs/1009.5961}{{\blue arXiv:1009.5961 [hep-ph]}}.

\bibitem{Kuchimanchi:2018ebf}
R.~Kuchimanchi, ``{Strong CP solution with soft PQ breaking}'',
\href{http://arxiv.org/abs/1805.00926}{{\blue arXiv:1805.00926 [hep-ph]}}.

\bibitem{Mimura:2019yfi}
Y.~Mimura, R.~N. Mohapatra, and M.~Severson, ``{Grand unified parity solution
  to the strong CP problem}'',
  \href{http://dx.doi.org/10.1103/PhysRevD.99.115025}{\blue Phys. Rev.
  {\bfseries D99} no.~11, (2019) 115025},
\href{http://arxiv.org/abs/1903.07506}{{\blue arXiv:1903.07506 [hep-ph]}}.

\bibitem{An:2009zh}
H.~An, X.~Ji, and F.~Xu, ``{P-odd and CP-odd Four-Quark Contributions to
  Neutron EDM}'', \href{http://dx.doi.org/10.1007/JHEP02(2010)043}{\blue JHEP
  {\bfseries 02} (2010) 043},
\href{http://arxiv.org/abs/0908.2420}{{\blue arXiv:0908.2420 [hep-ph]}}.

\bibitem{Khatsimovsky:1987fr}
V.~M. Khatsimovsky, I.~B. Khriplovich, and A.~S. Yelkhovsky, ``{Neutron
  Electric Dipole Moment, $T$ Odd Nuclear Forces and Nature of {CP}
  Violation}'',
\href{http://dx.doi.org/10.1016/S0003-4916(88)80015-0}{\blue Annals Phys.
  {\bfseries 186} (1988) 1--14}.

\bibitem{Xu:2009nt}
F.~Xu, H.~An, and X.~Ji, ``{Neutron Electric Dipole Moment Constraint on Scale
  of Minimal Left-Right Symmetric Model}'',
  \href{http://dx.doi.org/10.1007/JHEP03(2010)088}{\blue JHEP {\bfseries 03}
  (2010) 088},
\href{http://arxiv.org/abs/0910.2265}{{\blue arXiv:0910.2265 [hep-ph]}}.

\bibitem{Hisano:2012cc}
J.~Hisano, K.~Tsumura, and M.~J.~S. Yang, ``{QCD Corrections to Neutron
  Electric Dipole Moment from Dimension-six Four-Quark Operators}'',
  \href{http://dx.doi.org/10.1016/j.physletb.2012.06.038}{\blue Phys. Lett.
  {\bfseries B713} (2012) 473--480},
\href{http://arxiv.org/abs/1205.2212}{{\blue arXiv:1205.2212 [hep-ph]}}.

\bibitem{Braaten:1990gq}
E.~Braaten, C.-S. Li, and T.-C. Yuan, ``{The Evolution of Weinberg's Gluonic
  {CP} Violation Operator}'',
\href{http://dx.doi.org/10.1103/PhysRevLett.64.1709}{\blue Phys. Rev. Lett.
  {\bfseries 64} (1990) 1709}.

\bibitem{Chang:1991ry}
D.~Chang, T.~W. Kephart, W.-Y. Keung, and T.~C. Yuan, ``{The Chromoelectric
  dipole moment of the heavy quark and purely gluonic CP violating
  operators}'',
\href{http://dx.doi.org/10.1103/PhysRevLett.68.439}{\blue Phys. Rev. Lett.
  {\bfseries 68} (1992) 439--442}.

\bibitem{Chang:1992vs}
D.~Chang, T.~W. Kephart, W.-Y. Keung, and T.~C. Yuan, ``{An Effective field
  theory for the neutron electric dipole moment}'',
\href{http://dx.doi.org/10.1016/0550-3213(92)90465-N}{\blue Nucl. Phys.
  {\bfseries B384} (1992) 147--167}.

\bibitem{Brod:2018pli}
J.~Brod and E.~Stamou, ``{Electric dipole moment constraints on CP-violating
  heavy-quark Yukawas at next-to-leading order}'',
\href{http://arxiv.org/abs/1810.12303}{{\blue arXiv:1810.12303 [hep-ph]}}.

\bibitem{Hisano:2012sc}
J.~Hisano, J.~Y. Lee, N.~Nagata, and Y.~Shimizu, ``{Reevaluation of Neutron
  Electric Dipole Moment with QCD Sum Rules}'',
  \href{http://dx.doi.org/10.1103/PhysRevD.85.114044}{\blue Phys. Rev.
  {\bfseries D85} (2012) 114044},
\href{http://arxiv.org/abs/1204.2653}{{\blue arXiv:1204.2653 [hep-ph]}}.

\bibitem{Fuyuto:2012yf}
K.~Fuyuto, J.~Hisano, and N.~Nagata, ``{Neutron electric dipole moment induced
  by strangeness revisited}'',
  \href{http://dx.doi.org/10.1103/PhysRevD.87.054018}{\blue Phys. Rev.
  {\bfseries D87} no.~5, (2013) 054018},
\href{http://arxiv.org/abs/1211.5228}{{\blue arXiv:1211.5228 [hep-ph]}}.

\bibitem{Yamanaka:2018uud}
{ JLQCD}, N.~Yamanaka, S.~Hashimoto, T.~Kaneko, and H.~Ohki, ``{Nucleon charges
  with dynamical overlap fermions}'',
  \href{http://dx.doi.org/10.1103/PhysRevD.98.054516}{\blue Phys. Rev.
  {\bfseries D98} no.~5, (2018) 054516},
\href{http://arxiv.org/abs/1805.10507}{{\blue arXiv:1805.10507 [hep-lat]}}.

\bibitem{Shifman:1979if}
M.~A. Shifman, A.~I. Vainshtein, and V.~I. Zakharov, ``{Can Confinement Ensure
  Natural CP Invariance of Strong Interactions?}'',
\href{http://dx.doi.org/10.1016/0550-3213(80)90209-6}{\blue Nucl. Phys.
  {\bfseries B166} (1980) 493--506}.

\bibitem{Pospelov:2000bw}
M.~Pospelov and A.~Ritz, ``{Neutron EDM from electric and chromoelectric dipole
  moments of quarks}'',
  \href{http://dx.doi.org/10.1103/PhysRevD.63.073015}{\blue Phys. Rev.
  {\bfseries D63} (2001) 073015},
\href{http://arxiv.org/abs/hep-ph/0010037}{{\blue arXiv:hep-ph/0010037
  [hep-ph]}}.

\bibitem{Crewther:1979pi}
R.~J. Crewther, P.~Di~Vecchia, G.~Veneziano, and E.~Witten, ``{Chiral Estimate
  of the Electric Dipole Moment of the Neutron in Quantum Chromodynamics}'',
  \href{http://dx.doi.org/10.1016/0370-2693(80)91025-4,
  10.1016/0370-2693(79)90128-X}{\blue Phys. Lett. {\bfseries 88B} (1979) 123}.
[Erratum: Phys. Lett.91B,487(1980)].

\bibitem{Pich:1991fq}
A.~Pich and E.~de~Rafael, ``{Strong CP violation in an effective chiral
  Lagrangian approach}'',
\href{http://dx.doi.org/10.1016/0550-3213(91)90019-T}{\blue Nucl. Phys.
  {\bfseries B367} (1991) 313--333}.

\bibitem{Pospelov:1999ha}
M.~Pospelov and A.~Ritz, ``{Theta induced electric dipole moment of the neutron
  via QCD sum rules}'',
  \href{http://dx.doi.org/10.1103/PhysRevLett.83.2526}{\blue Phys. Rev. Lett.
  {\bfseries 83} (1999) 2526--2529},
\href{http://arxiv.org/abs/hep-ph/9904483}{{\blue arXiv:hep-ph/9904483
  [hep-ph]}}.

\bibitem{Pospelov:1999mv}
M.~Pospelov and A.~Ritz, ``{Theta vacua, QCD sum rules, and the neutron
  electric dipole moment}'',
  \href{http://dx.doi.org/10.1016/S0550-3213(99)00817-2}{\blue Nucl. Phys.
  {\bfseries B573} (2000) 177--200},
\href{http://arxiv.org/abs/hep-ph/9908508}{{\blue arXiv:hep-ph/9908508
  [hep-ph]}}.

\bibitem{Shindler:2015aqa}
A.~Shindler, T.~Luu, and J.~de~Vries, ``{Nucleon electric dipole moment with
  the gradient flow: The ?-term contribution}'',
  \href{http://dx.doi.org/10.1103/PhysRevD.92.094518}{\blue Phys. Rev.
  {\bfseries D92} no.~9, (2015) 094518},
\href{http://arxiv.org/abs/1507.02343}{{\blue arXiv:1507.02343 [hep-lat]}}.

\bibitem{Guo:2015tla}
F.~K. Guo, R.~Horsley, U.~G. Meissner, Y.~Nakamura, H.~Perlt, P.~E.~L. Rakow,
  G.~Schierholz, A.~Schiller, and J.~M. Zanotti, ``{The electric dipole moment
  of the neutron from 2+1 flavor lattice QCD}'',
  \href{http://dx.doi.org/10.1103/PhysRevLett.115.062001}{\blue Phys. Rev.
  Lett. {\bfseries 115} no.~6, (2015) 062001},
\href{http://arxiv.org/abs/1502.02295}{{\blue arXiv:1502.02295 [hep-lat]}}.

\bibitem{Chupp:2017rkp}
T.~Chupp, P.~Fierlinger, M.~Ramsey-Musolf, and J.~Singh, ``{Electric dipole
  moments of atoms, molecules, nuclei, and particles}'',
  \href{http://dx.doi.org/10.1103/RevModPhys.91.015001}{\blue Rev. Mod. Phys.
  {\bfseries 91} no.~1, (2019) 015001},
\href{http://arxiv.org/abs/1710.02504}{{\blue arXiv:1710.02504
  [physics.atom-ph]}}.

\bibitem{Demir:2002gg}
D.~A. Demir, M.~Pospelov, and A.~Ritz, ``{Hadronic EDMs, the Weinberg operator,
  and light gluinos}'',
  \href{http://dx.doi.org/10.1103/PhysRevD.67.015007}{\blue Phys. Rev.
  {\bfseries D67} (2003) 015007},
\href{http://arxiv.org/abs/hep-ph/0208257}{{\blue arXiv:hep-ph/0208257
  [hep-ph]}}.

\bibitem{He:1992db}
X.-G. He and B.~McKellar, ``{Large contribution to the neutron electric dipole
  moment from a dimension-six four quark operator}'',
\href{http://dx.doi.org/10.1103/PhysRevD.47.4055}{\blue Phys. Rev. {\bfseries
  D47} (1993) 4055--4058}.

\bibitem{Ottnad:2009jw}
K.~Ottnad, B.~Kubis, U.~G. Meissner, and F.~K. Guo, ``{New insights into the
  neutron electric dipole moment}'',
  \href{http://dx.doi.org/10.1016/j.physletb.2010.03.005}{\blue Phys. Lett.
  {\bfseries B687} (2010) 42--47},
\href{http://arxiv.org/abs/0911.3981}{{\blue arXiv:0911.3981 [hep-ph]}}.

\bibitem{Guo:2012vf}
F.-K. Guo and U.-G. Meissner, ``{Baryon electric dipole moments from strong CP
  violation}'', \href{http://dx.doi.org/10.1007/JHEP12(2012)097}{\blue JHEP
  {\bfseries 12} (2012) 097},
\href{http://arxiv.org/abs/1210.5887}{{\blue arXiv:1210.5887 [hep-ph]}}.

\bibitem{Seng:2014pba}
C.-Y. Seng, J.~de~Vries, E.~Mereghetti, H.~H. Patel, and M.~Ramsey-Musolf,
  ``{Nucleon electric dipole moments and the isovector parity- and
  time-reversal-odd pion-nucleon coupling}'',
  \href{http://dx.doi.org/10.1016/j.physletb.2014.07.014}{\blue Phys. Lett.
  {\bfseries B736} (2014) 147--153},
\href{http://arxiv.org/abs/1401.5366}{{\blue arXiv:1401.5366 [nucl-th]}}.

\bibitem{Jenkins:1991es}
E.~E. Jenkins and A.~V. Manohar, ``{Chiral corrections to the baryon axial
  currents}'',
\href{http://dx.doi.org/10.1016/0370-2693(91)90840-M}{\blue Phys. Lett.
  {\bfseries B259} (1991) 353--358}.

\bibitem{Fuchs:2003qc}
T.~Fuchs, J.~Gegelia, G.~Japaridze, and S.~Scherer, ``{Renormalization of
  relativistic baryon chiral perturbation theory and power counting}'',
  \href{http://dx.doi.org/10.1103/PhysRevD.68.056005}{\blue Phys. Rev.
  {\bfseries D68} (2003) 056005},
\href{http://arxiv.org/abs/hep-ph/0302117}{{\blue arXiv:hep-ph/0302117
  [hep-ph]}}.

\bibitem{Yamanaka:2017mef}
N.~Yamanaka, B.~K. Sahoo, N.~Yoshinaga, T.~Sato, K.~Asahi, and B.~P. Das,
  ``{Probing exotic phenomena at the interface of nuclear and particle physics
  with the electric dipole moments of diamagnetic atoms: A unique window to
  hadronic and semi-leptonic CP violation}'',
  \href{http://dx.doi.org/10.1140/epja/i2017-12237-2}{\blue Eur. Phys. J.
  {\bfseries A53} no.~3, (2017) 54},
\href{http://arxiv.org/abs/1703.01570}{{\blue arXiv:1703.01570 [hep-ph]}}.

\bibitem{Maiezza:2015lza}
A.~Maiezza, M.~Nemevšek, and F.~Nesti, ``{Lepton Number Violation in Higgs
  Decay at LHC}'',
  \href{http://dx.doi.org/10.1103/PhysRevLett.115.081802}{\blue Phys. Rev.
  Lett. {\bfseries 115} (2015) 081802},
\href{http://arxiv.org/abs/1503.06834}{{\blue arXiv:1503.06834 [hep-ph]}}.

\bibitem{Ruiz:2017nip}
R.~Ruiz, ``{Lepton Number Violation at Colliders from Kinematically
  Inaccessible Gauge Bosons}'',
  \href{http://dx.doi.org/10.1140/epjc/s10052-017-4950-2}{\blue Eur. Phys. J.
  {\bfseries C77} no.~6, (2017) 375},
\href{http://arxiv.org/abs/1703.04669}{{\blue arXiv:1703.04669 [hep-ph]}}.

\bibitem{Inami:1980fz}
T.~Inami and C.~S. Lim, ``{Effects of Superheavy Quarks and Leptons in
  Low-Energy Weak Processes k(L) to mu anti-mu, K+ to pi+ Neutrino
  anti-neutrino and K0 - anti-K0}'',
  \href{http://dx.doi.org/10.1143/PTP.65.297}{\blue Prog. Theor. Phys.
  {\bfseries 65} (1981) 297}.
[Erratum: Prog. Theor. Phys.65,1772(1981)].

\bibitem{Cho:1993zb}
P.~L. Cho and M.~Misiak, ``{b to s gamma decay in SU(2)-L x SU(2)-R x U(1)
  extensions of the Standard Model}'',
  \href{http://dx.doi.org/10.1103/PhysRevD.49.5894}{\blue Phys. Rev. {\bfseries
  D49} (1994) 5894--5903},
\href{http://arxiv.org/abs/hep-ph/9310332}{{\blue arXiv:hep-ph/9310332
  [hep-ph]}}.

\bibitem{Buras:1998raa}
A.~J. Buras, ``{Weak Hamiltonian, CP violation and rare decays}'', in {\em
  {Probing the standard model of particle interactions. Proceedings, Summer
  School in Theoretical Physics, NATO Advanced Study Institute, 68th session,
  Les Houches, France, July 28-September 5, 1997. Pt. 1, 2}}, pp.~281--539.
\newblock 1998.
\newblock
\href{http://arxiv.org/abs/hep-ph/9806471}{{\blue arXiv:hep-ph/9806471
  [hep-ph]}}.
\newblock

\bibitem{Misiak:1994zw}
M.~Misiak and M.~Munz, ``{Two loop mixing of dimension five flavor changing
  operators}'', \href{http://dx.doi.org/10.1016/0370-2693(94)01553-O}{\blue
  Phys. Lett. {\bfseries B344} (1995) 308--318},
\href{http://arxiv.org/abs/hep-ph/9409454}{{\blue arXiv:hep-ph/9409454
  [hep-ph]}}.

\bibitem{Degrassi:2005zd}
G.~Degrassi, E.~Franco, S.~Marchetti, and L.~Silvestrini, ``{QCD corrections to
  the electric dipole moment of the neutron in the MSSM}'',
  \href{http://dx.doi.org/10.1088/1126-6708/2005/11/044}{\blue JHEP {\bfseries
  11} (2005) 044},
\href{http://arxiv.org/abs/hep-ph/0510137}{{\blue arXiv:hep-ph/0510137
  [hep-ph]}}.

\bibitem{Scherer:2012xha}
S.~Scherer and M.~R. Schindler, ``{A Primer for Chiral Perturbation Theory}'',
\href{http://dx.doi.org/10.1007/978-3-642-19254-8}{\blue Lect. Notes Phys.
  {\bfseries 830} (2012) pp.1--338}.

\end{thebibliography}\endgroup

\end{document}